\documentclass[sigconf]{acmart}

\usepackage[utf8]{inputenc}
\usepackage{booktabs,multirow}

\usepackage{dsfont}
\usepackage{bbm}
\usepackage{arydshln}
\usepackage{hyperref}

\hypersetup{
    colorlinks=true,
    linkcolor=blue,
    filecolor=magenta,      
    urlcolor=cyan,
    }

\usepackage{color}
\usepackage{rotating}
\usepackage{amsmath}
\usepackage{array}
\usepackage{enumitem}
\usepackage[linesnumbered,algoruled,boxed,lined]{algorithm2e}
\usepackage{kotex}
\usepackage{balance}
\usepackage[short]{optidef}
\usepackage{graphicx}
\usepackage{dsfont}
\usepackage{threeparttable}
\usepackage{amsmath}
\usepackage{pifont}
\usepackage{adjustbox}

\usepackage{contour}
\usepackage[normalem]{ulem}

\contourlength{0.8pt}

\newcolumntype{P}[1]{>{\centering\arraybackslash}p{#1}}
\newcolumntype{R}[1]{>{\RaggedLeft\arraybackslash}p{#1}}

\newcommand{\ie}{{\it i.e.}}
\newcommand{\eg}{{\it e.g.}}

{}

\newcommand{\cmark}{\ding{51}}

\copyrightyear{2025}
\acmYear{2025}
\setcopyright{acmlicensed}
\acmConference[WSDM '25] {Proceedings of the Eighteenth ACM International Conference on Web Search and Data Mining}{March 10--14, 2025}{Hannover, Germany.}
\acmBooktitle{Proceedings of the Eighteenth ACM International Conference on Web Search and Data Mining (WSDM '25), March 10--14, 2025, Hannover, Germany}
\acmISBN{979-8-4007-1329-3/25/03}
\acmDOI{10.1145/3701551.3703554}

\begin{document}

\title{Temporal Linear Item-Item Model for Sequential Recommendation}

\author{Seongmin Park} \authornote{Both authors contributed equally to this research.}
\affiliation{
  \institution{Sungkyunkwan University}
  \city{Suwon}
  \country{Republic of Korea}}
\email{psm1206@skku.edu}

\author{Mincheol Yoon} \authornotemark[1]
\affiliation{
  \institution{Sungkyunkwan University}
  \city{Suwon}
  \country{Republic of Korea}}
\email{yoon56@skku.edu}

\author{Minjin Choi}
\affiliation{
  \institution{Sungkyunkwan University}
  \city{Suwon}
  \country{Republic of Korea}}
\email{zxcvxd@skku.edu}

\author{Jongwuk Lee}\authornote{Corresponding author}
\affiliation{
  \institution{Sungkyunkwan University}
  \city{Suwon}
  \country{Republic of Korea}}
\email{jongwuklee@skku.edu}



\begin{abstract}

In sequential recommendation (SR), neural models have been actively explored due to their remarkable performance, but they suffer from inefficiency inherent to their complexity. On the other hand, linear SR models exhibit high efficiency and achieve competitive or superior accuracy compared to neural models. However, they solely deal with the sequential order of items (\ie, \emph{sequential information}) and overlook the actual timestamp (\ie, \emph{temporal information}). It is limited to effectively capturing various \emph{user preference drifts} over time. To address this issue, we propose a novel linear SR model, named \emph{\textbf{T}empor\textbf{A}l \textbf{L}in\textbf{E}ar item-item model (\textbf{TALE})}, incorporating temporal information while preserving training/inference efficiency, with three key components. (i) \emph{Single-target augmentation} concentrates on a single target item, enabling us to learn the temporal correlation for the target item. (ii) \emph{Time interval-aware weighting} utilizes the actual timestamp to discern the item correlation depending on time intervals. (iii) \emph{Trend-aware normalization} reflects the dynamic shift of item popularity over time. Our empirical studies show that TALE outperforms ten competing SR models by up to 18.71\% gains on five benchmark datasets. It also exhibits remarkable effectiveness in evaluating long-tail items by up to 30.45\% gains. The source code is available at \url{https://github.com/psm1206/TALE}.

\end{abstract}

\maketitle


\section{Introduction}\label{sec:introduction}
Sequential recommendation (SR) aims to predict the next user interaction based on the user’s historical behavior in chronological order. Most existing studies focus on learning the transition pattern and correlation between items to capture hidden user preference drifts. Representative SR models~\cite{YueWHZMW24, HidasiKBT15GRU4Rec, KangM18SASRec, SunLWPLOJ19BERT4Rec, QiuHYW22, ZhouYZW22, DuYZQZ0LS23, ShinCWP24, WuT0WXT19} utilize neural architectures, such as recurrent neural networks (RNNs), graph neural networks (GNNs), and transformers. Because the transformer architecture~\cite{VaswaniSPUJGKP17} is suitable for learning long-term correlations and dependencies between items, transformer-based SR models~\cite{KangM18SASRec, SunLWPLOJ19BERT4Rec, QiuHYW22, ZhouYZW22, DuYZQZ0LS23, ShinCWP24} have shown outstanding performance.

However, existing SR models still struggle to address two challenges.
\textbf{(C1) Efficiency}: Neural SR models incur slow training and inference time. In particular, the time complexity of transformers is quadratic for the length of the item sequences.
\textbf{(C2) Temporal information}: Sequential information assumes that all time intervals are the same. In practice, the time interval between two consecutive items varies significantly. This highlights the need to handle temporal information in sequential item modeling.

\begin{figure}
\includegraphics[clip, trim=0cm 0cm 0cm 0cm, width=\columnwidth]{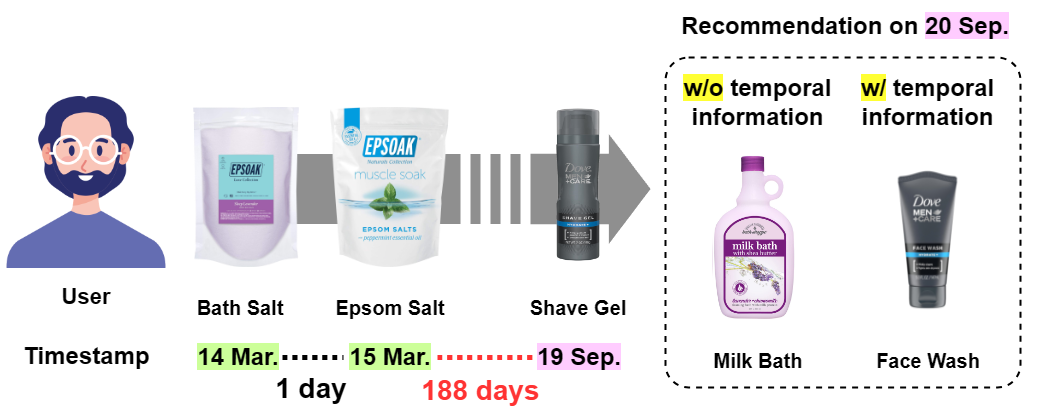} 
\vspace{-6mm}
\caption{An example of the user sequence with different time intervals on the Beauty dataset. }\label{fig:real_world_example}
\vspace{-4mm}

\end{figure}


To tackle the efficiency issue, recent studies~\cite{LiuCZZGWLFWHLL23, GaoZLZWGLY24, YueWHZMW24} have proposed to incorporate efficient mechanisms for SR. Specifically, LinRec~\cite{LiuCZZGWLFWHLL23} adopted L2-normalized linear attention in the self-attention block, improving the efficiency while preserving the capabilities of the conventional dot-product attention. LRURec~\cite{YueWHZMW24} utilized a linear recurrent unit to support recursive parallel training. Conversely, some studies~\cite{Steck19EASE, Steck20edlae, abs-1904-13033, MoonKL23RDLAE, ChoiKLSL21} proposed linear models without using complicated neural architectures. Among them, SLIST~\cite{ChoiKLSL21} is the representative linear SR model using item correlations and dependencies. Despite its simplicity, its accuracy is comparable to neural SR models, and it achieves efficient model training and inference. However, these efficient SR models only consider sequential information and do not account for temporal information.

Learning solely from sequential information without temporal dynamics does not represent \emph{user preference drifts} accurately. Figure~\ref{fig:real_world_example} provides an intuitive example of how recommendations vary depending on whether temporal information is taken into account.
Without temporal information, the SR model recommends ``\textit{Milk Bath}'' because the sequence is associated with two items related to the target item. On the other hand, the SR model using temporal information leverages the fact that it has been a while since ``\textit{Bath Salt}'' and \textit{Epsom Salt}'' were purchased to recommend ``\textit{Face Wash}'', which is related to the recently purchased ``\textit{Shave Gel}''.

\begin{figure}
\begin{tabular}{cc}
\includegraphics[height=3.0cm]{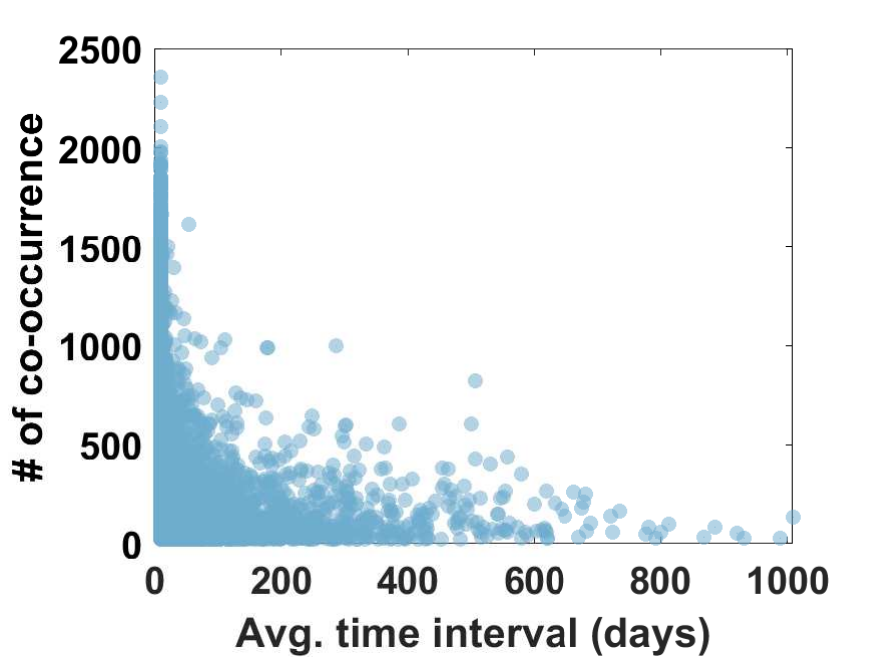} &
\includegraphics[height=3.0cm]{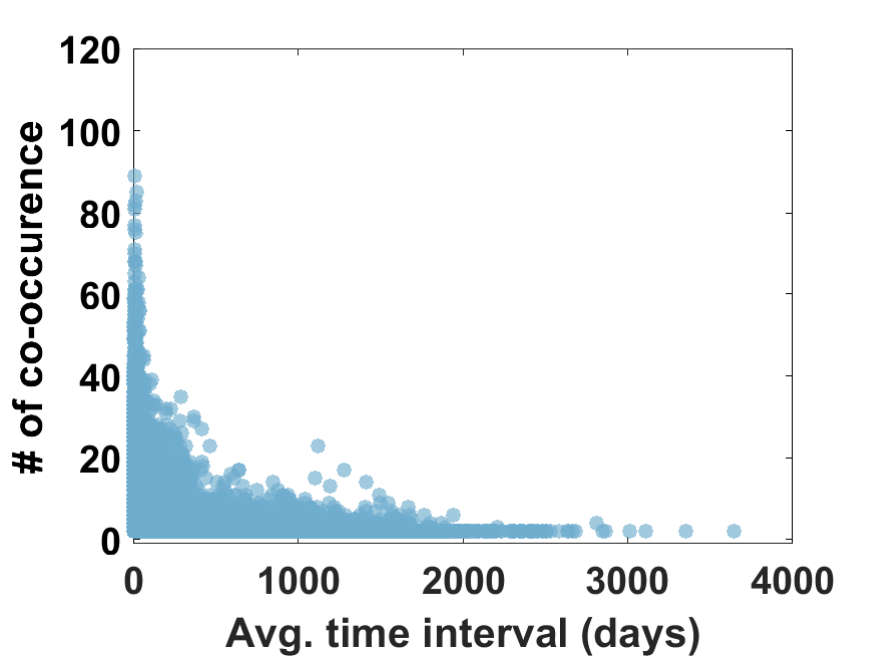} \\
\multicolumn{1}{c}{(a) ML-1M} &\multicolumn{1}{c}{(b) Beauty}
\end{tabular}

\vspace{-3mm}
\caption{Co-occurrences between two consecutive items over average time intervals on the ML-1M and Beauty datasets. Appendix~\ref{appen:intv_co_occur} shows a similar trend on Toys, Sports, and Yelp.}\label{fig:intv_co_occur}
\vspace{-3.5mm}
\end{figure}

It indicates that temporal information using actual timestamps provides a beneficial clue for capturing user preference drifts. Figure~\ref{fig:intv_co_occur} shows the co-occurrence of two consecutive items over various time intervals in real-world datasets. In the Beauty dataset, approximately 50\% of the consecutive item pairs occurred at intervals of more than 1 day, meaning that the temporal information between items varies significantly. In Appendix~\ref{appen:transition_prob}, we also show the existence of user preference drifts by utilizing item attributes (\eg, genre and item category) instead of co-occurrences.

Recent studies~\cite{LiWM20, TianLBWZ22, DangYGJ0XSL23, ZhangCWLX24} have attempted to combine temporal information with transformer-based or contrastive learning-based SR models. Specifically, TiSASRec~\cite{LiWM20} incorporated time interval-aware embedding vectors representing temporal context into the self-attention block. TiCoSeRec~\cite{DangYGJ0XSL23} introduced sequential augmentation methods using the time interval property. Besides, TCPSRec~\cite{TianLBWZ22} divided an interaction sequence into coherent subsequences to deal with temporal correlations among items. TGCL4SR~\cite{ZhangCWLX24} proposed local interaction sequences and global temporal graphs to capture item correlations. Although these temporal SR models improve recommendation performance, they do not address the efficiency issue.

To tackle the above two challenges (\ie, efficiency and temporal information), we propose an efficient linear model incorporating temporal information in SR, called \emph{\textbf{T}empor\textbf{A}l \textbf{L}in\textbf{E}ar item-item model (\textbf{TALE})}. We introduce three key components used in TALE. We first employ \emph{single-target augmentation} to ensure that temporal information is utilized without distortion. Subsequently, \emph{time interval-aware weighting} enables the linear model to harness temporal information in its recommendations. Lastly, to address the popularity bias problem~\cite{0007D0F0023}, which is a pervasive issue in recommender systems, we devise \emph{trend-aware normalization} considering the temporal context. We explain each component as follows.
\begin{itemize}[leftmargin=5mm]
    \item \textbf{Single-target augmentation}: While linear SR models typically leverage source and target matrices for training, we construct the target matrix using a single-item representation rather than multiple items. We prove that although previous models employing multiple target items strengthen long-interval collaborative signals, our single-target augmentation effectively captures temporal correlations between multiple source items and a single target item.
    

\vspace{0.5mm}
    \item \textbf{Time interval-aware weighting}: To directly incorporate temporal information into the linear model, we utilize actual timestamps to adjust the importance of items within the source items. In particular, this weighting strategy is divided into short and long time intervals. In this way, the item correlations with different time intervals can be discerned effectively.
    
\vspace{0.5mm}
    \item \textbf{Trend-aware normalization}: To reflect dynamic item popularity, we adopt a normalization method for eliminating popularity bias~\cite{0007D0F0023}. Although recent studies~\cite{WLCZDWSLW22, Wang0WFC19NGCF, 0001DWLZ020LightGCN, CaiHXR23, YuY00CN22} use various normalization methods, they fail to account for temporal context since they use global popularity over the entire period\footnote{For example, applying normalization to the Amazon 2014 dataset would use the accumulated popularity over 14 years, so it fails to consider item trends that can change several times within a single year.}. We utilize local item popularity in the recent $N$ days, reflecting the dynamic changes in item popularity.

\end{itemize}

We extensively evaluate the effectiveness of TALE by comparing it with ten existing SR models on five benchmark datasets, achieving up to 18.71\% higher accuracy and 6.9x to 57x faster training times. Notably, TALE shows up to 30.45\% performance gains on long-tail items, alleviating popularity bias.




\section{Background}\label{sec:background}

\subsection{Problem Statement}

Given a historical user sequence $\mathcal{S}^u=[i_1^u, i_2^u, ..., i_{|\mathcal{S}^u|}^u]$, SR aims to recommend the next item that user $u$ is likely to interact with. The timestamp sequence is denoted by $\mathcal{T}^u=[t_1^u, t_2^u, ..., t_{|\mathcal{S}^u|}^u]$, indicating the temporal information corresponding to $\mathcal{S}^u$. The entire training sequence is stacked as a set of users $\mathcal{U} \in \mathbb{R}^{m}$, and each item in the sequences belongs to a set of items $\mathcal{I} \in \mathbb{R}^{n}$.


The SR model learns an optimal model parameter ${\Theta}^*$ to predict user $u$'s ground-truth item gt($u$) given user $u$'s item sequence $\mathcal{S}^u$. Temporal SR models further exploit the timestamp sequence $\mathcal{T}^u$ for model training.
\begin{equation}
{\Theta}^* = \underset{\mathbf{\Theta}}{\text{argmax}} \sum_{u=1}^{m} p(i_{|\mathcal{S}^u|+1}^u=\text{gt}(u) | \mathcal{S}^u, \mathcal{T}^u, \Theta).
\end{equation}
In this process, the core part of temporal SR models is \emph{how to incorporate temporal information into the model training}.


\subsection{Linear Item-Item Models}
Given a set $\mathcal{S}$ of user-item interactions, we derive two interaction matrices $\mathbf{X} \in \{0,1\}^{m \times n}$ and $\mathbf{Y} \in \{0,1\}^{m \times n}$, corresponding to training input and target matrices, respectively. Let $\mathbf{X}_{u,j}$ be 1 if user $u$ and item $j$ have interacted, and 0 otherwise. Similarly, $\mathbf{Y}_{u,j}$ is 1 for items to target and 0 otherwise.
Given sequence-item interaction matrices $\mathbf{X}$ and $\mathbf{Y}$, linear item-item models learn an item-to-item weight matrix $\mathbf{B} \in \mathbb{R}^{n \times n}$.
\begin{equation}\label{eq:lae_obj}
  \min_{\mathbf{B}} \|\mathbf{Y} - \mathbf{X} \mathbf{B}\|_F^2 + \lambda \|\mathbf{B}\|_F^2,
\end{equation}
where $\lambda$ is the hyperparameter to adjust the regularization in $\mathbf{B}$. In Section~\ref{sec:model}, we will present how to determine the interaction matrices to handle temporal information in detail.

For inference, the prediction score $s_{u,j}$ between the user $u$ and the item $j$ is computed as follows.
\begin{equation}\label{eq:prediction}
    s_{u,j} = \mathbf{X}_{u,*} \cdot \hat{\mathbf{B}}_{*,j},
\end{equation}
where $\mathbf{X}_{u,*}$ is the row vector for the user $u$ in $\mathbf{X}$, and $\hat{\mathbf{B}}_{*,j}$ is the column vector for the item $j$ in $\hat{\mathbf{B}}$.

\subsection{Normalization}\label{sec:background_normalization}

As the inherent problem in recommendations, popularity bias~\cite{0007D0F0023} hurts user experience by predominantly yielding popular items to users. To eliminate the popularity bias, a lot of recommender models~\cite{GuptaGMVS2019, WangYMZCLM22, Chen23, KimPK23, WLCZDWSLW22, Wang0WFC19NGCF, 0001DWLZ020LightGCN, CaiHXR23, YuY00CN22, abs-1904-13033} utilize normalization. A few studies~\cite{GuptaGMVS2019, WangYMZCLM22, Chen23, KimPK23} apply normalization to user/item embeddings to adjust their magnitude, while others~\cite{WLCZDWSLW22, Wang0WFC19NGCF, 0001DWLZ020LightGCN, CaiHXR23, YuY00CN22} implement normalization on the adjacency matrix in GNN-based models to reduce the influence of popular nodes during message passing.

Meanwhile, linear models can utilize normalization to decrease the influence of popular items. It involves dividing the interaction matrix by the square root of the user/item popularity.
\begin{align} \label{eq:sym_norm}
    \Tilde{\mathbf{X}} = \mathbf{D}_U^{-1/2} \mathbf{X} \mathbf{D}_I^{-1/2} = \mathbf{W}_{\text{norm}} \odot \mathbf{X},  
\end{align}
where $\mathbf{D}_I = \text{diag} (p_{1}, p_{2},~...,~p_{n}) \in \mathbb{R}^{n \times n}$, and $p_{j}=\sum_{v=1}^{m} \mathbf{X}_{v,j}$ is the popularity of the item $i_j$. It can also be expressed as an interaction-wise operation, \ie, $\mathbf{W}_{\text{norm}} \in \mathbb{R}^{m \times n}$. By substituting Eq.~\eqref{eq:sym_norm} into the matrices $\mathbf{X}$ and $\mathbf{Y}$ of Eq.~\eqref{eq:lae_obj}, normalization can be applied for linear item-item models.

Existing normalization methods~\cite{WLCZDWSLW22, Wang0WFC19NGCF, 0001DWLZ020LightGCN, CaiHXR23, YuY00CN22} utilize long-term popularity over the entire time period, being limited to considering item trends. For instance, items that are popular only during specific periods (\ie, fad items) and items that have maintained decent attraction over a long period (\ie, steady sellers) can be normalized equally if they have the same long-term popularity. However, fad items should be considered popular or unpopular depending on the time. 



\section{Temporal Linear Item-Item Model}\label{sec:model}
\begin{figure*}
\includegraphics[clip, trim=0cm 0cm 0cm 0cm, width=\textwidth]{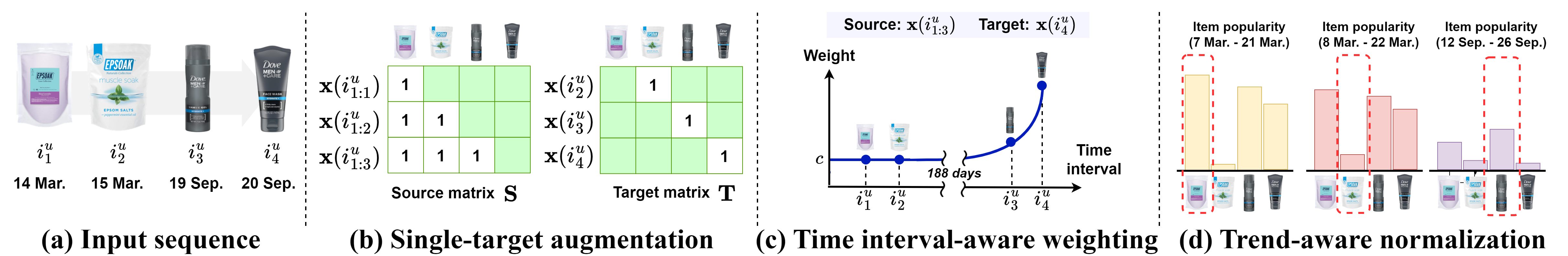}

\vspace{-2mm}
\caption{Illustration of three components of TALE. Single-target augmentation takes each item in the input sequence as the target and the previous items as the source to form the training matrices. Time interval-aware weighting adjusts the significance of each source item, and Trend-aware normalization normalizes items based on their popularity around the time.}\label{fig:framework}
\end{figure*}

In this section, we first present how sequential information is used in linear item-item models. We then propose \emph{\textbf{T}empor\textbf{A}l \textbf{L}in\textbf{E}ar Item-Item model (\textbf{TALE})}, which effectively incorporates temporal information into linear models in SR.

\vspace{1mm}
\noindent
\textbf{Sequential linear item-item models}.
Several studies~\cite{JannachL17, GargGMVS19, ChoiKLSL21} have exploited the efficiency of linear models in SR. Notably,~\citet{ChoiKLSL21} proposes the sequential linear item-item model, \ie, SLIST, which consists of two parts such as SLIS and SLIT. In this paper, we focus on improving SLIT because it mainly handles sequential information for linear models. Specifically, SLIT~\cite{ChoiKLSL21} utilizes two components to reflect sequential information as follows.

\begin{itemize}[leftmargin=5mm]
    \item \textbf{Multi-target augmentation}: It uses a source matrix $\mathbf{S}$ and a target matrix $\mathbf{T}$ instead of $\mathbf{X}$ and $\mathbf{Y}$ in Eq.~\eqref{eq:lae_obj}. Specifically, it splits an entire training matrix into $\mathbf{S}$ and $\mathbf{T}$ in disjoint sequential order. For example, if the user sequence is $[i_1^u, i_2^u, i_3^u, i_4^u]$, then (source, target) sub-sequence pairs are constructed by $([i_1^u], [i_2^u, i_3^u, i_4^u])$, $([i_1^u, i_2^u], [i_3^u, i_4^u])$, and $([i_1^u, i_2^u, i_3^u],$ $[i_4^u])$. Consequently, $\mathbf{S}$ and $\mathbf{T}$ are augmented with multiple sub-sequences. Using $\mathbf{S}$ and $\mathbf{T}$, SLIT then learns a transition matrix $\mathbf{B}$, representing the transition from source items to target items.
\vspace{0.5mm}
    \item \textbf{Position-aware weighting}: It is a weighting scheme using position gap outlined in~\cite{GargGMVS19}. This scheme prioritizes higher weights for items that are close to the last item in the source sequence and the first item in the target sequence. In other words, the relative order within the sequence reflects different sequential information.
\end{itemize}

Although SLIT~\cite{ChoiKLSL21} handles sequential information of items, it does not incorporate temporal information, thereby failing to reflect the user preference drift over various time intervals.

\subsection{Overall Framework}

As shown in Figure~\ref{fig:framework}, TALE consists of three key components to reflect temporal information. (i) \emph{Single-target augmentation} utilizes a single target item to prevent unnecessary intervention in learning the temporal correlation between input and target items. (ii) \emph{Time interval-aware weighting} discerns different time intervals using actual timestamps. (iii) \emph{Trend-aware normalization} reflects dynamic item trends and reduces the popularity bias of items in the sequence. 

Specifically, TALE learns the item-item transition matrix $\mathbf{\hat{B}_{\text{TALE}}}$ from source items to a single target item. By reformulating Eq.~\eqref{eq:lae_obj}, the objective function of TALE is as follows.
\begin{equation} \label{eq:tale_obj}
\begin{aligned}
  \min_{\mathbf{B}} \| \Tilde{\mathbf{T}} - & \Tilde{\mathbf{S}} \mathbf{B} \|_F^2 + \lambda \|\mathbf{B}\|_F^2, \\
  \text{where} \ \Tilde{\mathbf{T}} = \mathbf{W}_{\text{trend}} \odot \mathbf{T} \ & \text{and} \ \Tilde{\mathbf{S}} = \mathbf{W}_{\text{time}} \odot \mathbf{W}_{\text{trend}} \odot \mathbf{S}.
\end{aligned}
\end{equation}

Here, $\Tilde{\mathbf{T}}, \Tilde{\mathbf{S}} \in \mathbb{R}^{m' \times n}$ are target and source matrices for training TALE. Let $\mathbf{W}_{\text{time}}, \mathbf{W}_{\text{trend}} \in \mathbb{R}^{m' \times n}$ denote weight matrices for time interval-aware weighting and trend-aware normalization, and $\mathbf{W}_{\text{trend}}$ serves as the equivalent of $\mathbf{W}_{\text{norm}}$ in Eq.~\eqref{eq:sym_norm}. Let $m'$ be the number of all augmented sub-sequences in $\Tilde{\mathbf{T}}$ and $\Tilde{\mathbf{S}}$, and $\odot$ means the element-wise product. Note that computing $\Tilde{\mathbf{T}}$ and $\Tilde{\mathbf{S}}$ does not increase model training and inference costs.

The closed-form solution for Eq.~\eqref{eq:tale_obj} is derived through convex optimization, achieving fast model training.
\begin{align}\label{eq:tale_solution}
  \hat{\mathbf{B}}_{\text{TALE}} = \left({\Tilde{\mathbf{S}}}^{\top} \Tilde{\mathbf{S}} + \lambda \mathbf{I} \right)^{-1}  {\Tilde{\mathbf{S}}}^{\top} \Tilde{\mathbf{T}}.
\end{align}

For inference, TALE simply utilizes Eq.~\eqref{eq:prediction}, as used in \cite{GargGMVS19, ChoiKLSL21, Steck19EASE, Steck20edlae, MoonKL23RDLAE}. Unlike model training, it assigns item weights based on the relative order of test items. Because the temporal correlation between items is learned in $\hat{\mathbf{B}}_{\text{TALE}}$, it is sufficient to consider the relative order of items.



\subsection{Single-Target Augmentation} \label{sec:single_aug}

To overcome a limitation of multi-target augmentation~\cite{ChoiKLSL21}, we introduce \emph{single-target augmentation} to effectively incorporate temporal information into linear models (Figure~\ref{fig:framework}-(b)). It uses only the first target item per sub-sequence in the target matrix. For example, if user $u$ sequence is $[i_1^u, i_2^u, i_3^u, i_4^u]$, then (source, target) sequence pairs are augmented by $([i_1^u], [i_2^u])$, $([i_1^u, i_2^u], [i_3^u])$, and $([i_1^u, i_2^u, i_3^u], [i_4^u])$. As a result, it prevents unnecessary correlations between long-distance items.





\vspace{1mm}
\noindent
\textbf{Theoretical analysis}.
The objective function with multi-target augmentation is derived as follows, and the detailed proof is in Appendix~\ref{appen:proof_multi_target}. $\mathbf{T}^{\text{multi}}$ and $\mathbf{T}^{\text{single}}$ denote target matrix for multi- and single-target augmentation, respectively.
\begin{align} 
    \hat{\mathbf{B}}_{\text{multi}} 
    & =  \underset{\mathbf{B}}{\text{argmin}}\sum_{u'=1}^{m'} \|\mathbf{T}^{\text{multi}}_{u',*}-\mathbf{S}_{u',*}\mathbf{B}\|^2_F \\
    & = \underset{\mathbf{B}}{\text{argmin}}\sum_{u=1}^{m}\sum_{l=1}^{|\mathcal{S}^u|-1}\| \mathbf{x}(i_{l+1:|\mathcal{S}^u|}^u) - \mathbf{x}(i_{1:l}^u)\mathbf{B}\|^2_{F} \label{eq:multi_objective_1} \\
    & = \underset{\mathbf{B}}{\text{argmin}}\sum_{u=1}^{m}\sum_{s=1}^{|\mathcal{S}^u|-1}\sum_{h=s+1}^{|\mathcal{S}^u|} (h-s)\|\mathbf{x}(i_{h}^u) - \mathbf{x}(i_{s}^u)\mathbf{B}\|^2_{F} + \Tilde{\textbf{B}}^{S}, \label{eq:multi_objective_cvt}
\end{align}
where $\mathbf{x}(i_j^u) \in \mathbb{R}^{1 \times n}$ represents a one-hot vector of user $u$ with item $i_j^u$ is marked as 1, and $\mathbf{x}(i_{j:l}^u)$ denotes a multi-hot vector, marked 1 for an item sequence $[i_j^u, \dots, i_{l}^u]$. In Eq.~\eqref{eq:multi_objective_1}, $\mathbf{x}(i_{l+1:|\mathcal{S}^u|}^u)$ and $\mathbf{x}(i_{1:l}^u)$ mean the target and source sequence, respectively. Note that $\Tilde{\textbf{B}}^{S}$ refers to terms that are only affected by source sequences, and since they use the same source matrix, the effect of the terms can be ignored (The detailed form of $\Tilde{\textbf{B}}^{S}$ can be found in the Appendix~\ref{appen:proof_multi_target}.). 
From Eq.~\eqref{eq:multi_objective_cvt}, the objective function with multi-target augmentation gives a weight of $(h-s)$ (\ie, position gap) in learning the transition scores from the $s$-th item to the $h$-th item. Since $(h-s) \ge 1$, this inadvertently increases the weight between distant items. 

To solve this problem, we simply employ the next item as a target item. The position weight $(h-s)$ can be omitted in the objective function with single-target augmentation, so the learned item correlation is influenced by time interval-aware weighting. The detailed proof is presented in Appendix~\ref{appen:proof_single_target}.
\begin{align} 
    \hat{\mathbf{B}}_{\text{single}} 
    & = \underset{\mathbf{B}}{\text{argmin}}\sum_{u'=1}^{m'} \|\mathbf{T}^{\text{single}}_{u',*}-\mathbf{S}_{u',*}\mathbf{B}\|^2_F \\
    & = \underset{\mathbf{B}}{\text{argmin}}\sum_{u=1}^{m}\sum_{l=1}^{|\mathcal{S}^u|-1}\| \mathbf{x}(i_{l+1}^u)-\mathbf{x}(i_{1:l}^u)\mathbf{B}\|^2_{F} \label{eq:slit1target_objective_1} \\
    & = \underset{\mathbf{B}}{\text{argmin}}\sum_{u=1}^{m}\sum_{s=1}^{|\mathcal{S}^u|-1}\sum_{h=s+1}^{|\mathcal{S}^u|}\|\mathbf{x}(i_{h}^u) - \mathbf{x}(i_{s}^u)\mathbf{B}\|^2_{F} + \Tilde{\textbf{B}}^{S}. \label{eq:slit1target_objective_2}
\end{align}

We also empirically show that single-target augmentation, which does not emphasize the signal between distant items, performs better than multi-target augmentation (Refer to Section~\ref{sec:exp_analysis}). Also, it helps enhance training efficiency by reducing the computation of matrix multiplication with a sparser target matrix.


\subsection{Time Interval-aware Weighting} \label{sec:time_weight}
Owing to single-target augmentation, \emph{time interval-aware weighting} is capable of injecting temporal information into linear models as intended. As depicted in Figure~\ref{fig:framework}-(c), it utilizes actual timestamps to control item weights in the source matrix and capture temporal item correlations. Specifically, each element of time interval-aware weight matrix $\mathbf{W}_{\text{time}}$ is computed using weighting function $w_{\text{time}}(\cdot, \cdot)$, which calculates item weights using the time interval between each source item and the target item.
\begin{align} \label{eq:time_interval_weighting}
    w_{\text{time}}(\mathcal{S}^u_{k}, i_{s}^u) = \max \left(\exp\left(-\frac{t_{|\mathcal{S}^u_{k}|}^u - t_{s}^u} {\tau_{\text{time}}}\right), c \right),
\end{align}
where $\mathcal{S}^u_{k}$ denotes $k$-th augmented sequence for user $u$, and $t_{|\mathcal{S}^u_{k}|}^u$ and $t_{s}^u$ are timestamps of a target item $i_{|\mathcal{S}^u_{k}|}^u$ and a source item $i_{s}^u$. $c \in [0, 1]$ is a hyperparameter for the threshold, and $\tau_{\text{time}}$ adjusts the time decay in sequences.

The time interval-aware weight becomes larger for shorter time intervals and smaller for longer time intervals. To consider long-time dependency, we weigh items with long-time intervals by threshold $c$ to learn weak item relationships. For instance, if there is no threshold $c$, the weight of two items with a 30-day interval is as $e^{(-30 / \tau)} \approx 0$. However, if a user who has interacted in the past interacts again after a long time interval, it is still a weak collaborative signal from recent and past items. Therefore, the threshold $c$ is used to account for this long-time dependency, allowing TALE to capture a wider variety of item relationships.

\subsection{Trend-aware Normalization} \label{sec:trend_norm}


By harnessing temporal information through the two aforementioned methods, distant popular items with low correlation can be learned with lower weights, thereby indirectly mitigating popularity bias. However, since the popularity bias still persists, we introduce \emph{trend-aware normalization} that takes temporal context into account. (Appendix~\ref{appen:effect_trend_norm} shows the debiasing effect of our proposed components.)
It normalizes the training matrix using the item popularity of the recent $N$ days based on the interaction timestamps. For example, in Figure~\ref{fig:framework}-(d), items interacted in March (or September) are normalized using only March (or September) popularity. The trend-aware popularity matrix $\mathbf{P}_{\text{trend}} \in \mathbb{R}^{m' \times n}$ is formulated as follows\footnote{Trend-aware popularity incorporates both past and future information (both past $N$ days and future $N$ days). Since commonly used normalization methods~\cite{Wang0WFC19NGCF, 0001DWLZ020LightGCN, CaiHXR23} utilize both past and future information,  we designed trend-aware normalization to be a generalized form of them. Moreover, we observed that even when using only past information (past $2N$ days), it performs similarly to trend-aware normalization.}. The trend-aware normalization matrix $\mathbf{W}_{\text{trend}} = (\mathbf{P}_{\text{trend}})^{-\gamma}$ is used in Eq.~\eqref{eq:tale_obj}, with $\gamma$ set to $1/2$ as in existing work~\cite{0001DWLZ020LightGCN, Wang0WFC19NGCF, YuXCCHY24XSimGCL, YuY00CN22}.
\begin{equation}
\begin{aligned}
    \mathbf{P}_{\text{trend}} = 
    \begin{bmatrix} 
   p_{1,1}(N) & p_{1,2}(N) & \cdots & p_{1,n}(N)  \\
   p_{2,1}(N) & p_{2,2}(N) & \cdots & p_{2,n}(N)  \\
   \vdots  & \vdots  & \ddots & \vdots  \\
   p_{m',1}(N) & p_{m',2}(N) & \cdots & p_{m',n}(N)  \\
   \end{bmatrix}, \\
    \text{where} \ p_{u,j}(N) =\sum_{v=1}^{m} \mathds{1}( |t_{u,j} - t_{v,j}| \leq N).
\end{aligned}
\end{equation}

Let $t_{u,j}$ denote a timestamp of the interaction of user $u$ and item $j$, and $\mathds{1} (\cdot)$ is an indicator function.

Since $\mathbf{P}_{\text{trend}}$ can be pre-computed, there is no additional computation cost in the training/inference phase. We do not perform trend-aware normalization for users because there are fewer interactions per user compared to per item, resulting in less diversity in user trend patterns (Additionally, we found that user-side trend-aware normalization does not contribute to performance improvement.).

The cumulative popularity over the entire period can be thought of as a special case of trend-aware popularity. As the window size $N$ approaches infinity, they converge to the same popularity.
\begin{align} \label{eq:general_popularity}
    p_{j}= \lim_{N \rightarrow \infty} p_{*,j}(N).
\end{align}

\section{Experimental Setup}\label{sec:setup}

\begin{table}[t]
\begin{center}
\caption{Statistics of five benchmark datasets. `Avg. Interval' indicates the average of time intervals for consecutive items.}
\vspace{-2mm}
\label{tab:statistics}
\begin{tabular}{c|ccccc}
\toprule
Dataset & ML-1M & Beauty & Toys & Sports & Yelp  \\
\midrule
\#Users & 6,040 & 22,363 & 19,412 & 29,858 & 30,494 \\
\#Items & 3,953 & 12,101 & 11,924 & 18,357 & 20,061 \\
\#Inter. & 999,611 & 198,502 & 167,597 & 296,337 & 317,078 \\
Density & 5.19\% & 0.07\% & 0.07\% & 0.06\% & 0.05\% \\
Avg. Length & 165.6 & 8.9 & 8.6 & 8.3 & 10.4 \\
Avg. Interval & 0.6d & 69.6d & 86.0d & 74.1d & 18.6d \\
\bottomrule
\end{tabular}

\end{center}
\vspace{-1mm}
\end{table}

\begin{table*}[t]  
\caption{Accuracy comparison for the eight traditional SR models on five datasets. The best results are marked in \textbf{bold}, and the second-best results are \underline{underlined}. `Imp.' indicates the improvement ratio of TALE compared to SLIST~\cite{ChoiKLSL21}.}
\label{tab:overall_perform}
\vspace{-3mm}
\begin{center}
\renewcommand{\arraystretch}{0.95} 

\begin{tabular}{c|c|cccc|ccc|cc|c}
\toprule
Dataset & Metric & SASRec & DuoRec & FEARec & BSARec & TiSASRec & TCPSRec & TiCoSeRec & SLIST & TALE & Imp. \\
\midrule
\multirow{5}{*}{ML-1M} & HR@1 & 0.0611 & 0.0606 & 0.0594 & 
\underline{0.0623} & 0.0578 & 0.0540 & 0.0031 & 0.0596 & \textbf{0.0656} & 10.07\%  \\ 
         & HR@5 & 0.1763 & {0.1838} & 0.1829 & \textbf{0.1868} & 0.1851 & 0.1697 & 0.0884 & 0.1710 & \underline{0.1854} & 8.42\%  \\
         & HR@10 & {0.2697} & {0.2697} & 0.2672 & \textbf{0.2752} & \underline{0.2700} & 0.2632 & 0.1719 & 0.2373 & 0.2546 & 7.29\%  \\
         & NDCG@5 & 0.1197 & 0.1230 & 0.1220 & \underline{0.1261} & 0.1216 & 0.1124 & 0.0452 & 0.1164 & \textbf{0.1275} & 9.54\%  \\ 
         & NDCG@10 & 0.1496 & \underline{0.1506} & 0.1492 & \textbf{0.1543} & 0.1488 & 0.1423 & 0.0721 & 0.1377 & 0.1500 & 8.93\% \\
\midrule
\multirow{5}{*}{Beauty} & HR@1 & 0.0071 & 0.0122 & 0.0126 & 0.0112 & 0.0092 & 0.0080 & 0.0178 & \underline{0.0267} & \textbf{0.0288} & 7.87\%  \\ 
         & HR@5 & 0.0558 & 0.0566 & 0.0579 & 0.0553 & 0.0579 & 0.0567 & 0.0504 & \underline{0.0597} & \textbf{0.0674} & 12.90\%  \\ 
         & HR@10 & 0.0821 & 0.0864 & 0.0876 & \textbf{0.0888} & 0.0842 & 0.0855 & 0.0740 & 0.0797 & \underline{0.0887} & 11.29\%  \\ 
         & NDCG@5 & 0.0319 & 0.0350 & 0.0356 & 0.0338 & 0.0340 & 0.0332 & 0.0343 & \underline{0.0435} & \textbf{0.0485} & 11.49\%  \\ 
         & NDCG@10 & 0.0404 & 0.0446 & 0.0451 & 0.0445 & 0.0425 & 0.0424 & 0.0418 & \underline{0.0499} & \textbf{0.0554} & 11.02\%  \\ 
\midrule
\multirow{5}{*}{Toys} & HR@1 & 0.0059 & 0.0095 & 0.0107 & 0.0195 & 0.0094 & 0.0043 & 0.0176 & \underline{0.0333} & \textbf{0.0355} & 6.61\%  \\ 
         & HR@5 & 0.0601 & 0.0642 & 0.0667 & 0.0606 & 0.0675 & 0.0648 & 0.0545 & \underline{0.0722} & \textbf{0.0815} & 12.88\%  \\ 
         & HR@10 & 0.0863 & 0.0941 & \underline{0.0967} & 0.0906 & 0.0963 & 0.0936 & 0.0801 & 0.0919 & \textbf{0.1021} & 11.10\%  \\ 
         & NDCG@5 & 0.0342 & 0.0382 & 0.0401 & 0.0403 & 0.0392 & 0.0358 & 0.0363 & \underline{0.0535} & \textbf{0.0594} & 11.03\%  \\ 
         & NDCG@10 & 0.0426 & 0.0479 & 0.0497 & 0.0500 & 0.0485 & 0.0452 & 0.0445 & \underline{0.0599} & \textbf{0.0661} & 10.35\%  \\ 
\midrule
\multirow{5}{*}{Sports} & HR@1 & 0.0021 & 0.0053 & 0.0044 & 0.0022 & 0.0026 & 0.0018 & 0.0109 & \underline{0.0155} & \textbf{0.0184} & 18.71\%  \\ 
         & HR@5 & 0.0292 & 0.0331 & 0.0329 & 0.0324 & 0.0308 & 0.0331 & 0.0342 & \underline{0.0380} & \textbf{0.0418} & 10.00\%  \\ 
         & HR@10 & 0.0462 & 0.0498 & 0.0511 & \underline{0.0544} & 0.0473 & 0.0521 & 0.0515 & 0.0517 & \textbf{0.0564} & 9.09\%  \\ 
         & NDCG@5 & 0.0162 & 0.0195 & 0.0192 & 0.0175 & 0.0172 & 0.0180 & 0.0226 & \underline{0.0271} & \textbf{0.0305} & 12.55\%  \\ 
         & NDCG@10 & 0.0217 & 0.0249 & 0.0250 & 0.0245 & 0.0226 & 0.0241 & 0.0282 & \underline{0.0315} & \textbf{0.0352} & 11.75\%  \\ 
\midrule
\multirow{5}{*}{Yelp} & HR@1 & \underline{0.0219} & 0.0212 & 0.0212 & 0.0218 & 0.0218 & \textbf{0.0220} & 0.0065 & 0.0176 & \textbf{0.0220} & 25.00\%  \\ 
         & HR@5 & 0.0417 & 0.0437 & 0.0435 & 0.0468 & 0.0430 & 0.0461 & 0.0231 & \textbf{0.0563} & \underline{0.0558} & -0.89\%  \\ 
         & HR@10 & 0.0605 & 0.0622 & 0.0619 & 0.0680 & 0.0607 & 0.0657 & 0.0382 & \textbf{0.0720} & \underline{0.0707} & -1.81\%  \\ 
         & NDCG@5 & 0.0317 & 0.0327 & 0.0324 & 0.0345 & 0.0326 & 0.0342 & 0.0141 & \underline{0.0375} & \textbf{0.0394} & 5.07\%  \\ 
         & NDCG@10 & 0.0378 & 0.0386 & 0.0383 & 0.0413 & 0.0382 & 0.0404 & 0.0191 & \underline{0.0426} & \textbf{0.0442} & 3.76\% \\

\bottomrule
\end{tabular}

\vspace{-2mm}
\end{center}
\end{table*}

\noindent
\textbf{Datasets}. 
As shown in Table~\ref{tab:statistics}, we use five benchmark datasets, all of which are pre-processed with a 5-core setting, \ie, users and items with less than 5 interactions are removed~\cite{DuYZQZ0LS23, DangYGJ0XSL23, TianLBWZ22}.
\begin{itemize}[leftmargin=5mm]
    \item \textbf{ML-1M\footnote{\url{https://grouplens.org/datasets/movielens/1m/}}} is composed of a collection of users' movie reviews. It is about 100 times denser than other datasets, and the average length of user history is up to 20 times longer.
\vspace{0.5mm}
    \item \textbf{Beauty, Toys, and Sports\footnote{\url{https://cseweb.ucsd.edu/~jmcauley/datasets/amazon/links.html}}} are product review datasets with a total period of 14 years, which is quite long compared to other datasets. We used the widely used 2014 versions. 
\vspace{0.5mm}
    \item \textbf{Yelp\footnote{\url{https://www.yelp.com/dataset/}}} consists of user reviews for the businesses. Following previous studies~\cite{TianLBWZ22, DangYGJ0XSL23, DuYZQZ0LS23}, we used transaction records between January 1, 2019 and January 1, 2020. 
\end{itemize}

\vspace{1mm}
\noindent
\textbf{Evaluation protocol}. We adopt a \emph{leave-one-out strategy} following the convention~\cite{DuYZQZ0LS23, DangYGJ0XSL23, ShinCWP24}. For each user, we utilize the last item as the test item, the second last item as the validation item, and the remaining items for the training set. In the inference phase, we take training and validation items as inputs and evaluate the performance on the test item. For evaluation metrics, we adopt \emph{Hit Ratio (HR@$K$)} and \emph{Normalized Discounted Cumulative Gain (NDCG@$K$)}, with $K$ set to 1, 5, and 10 as default. According to item popularity of the training set, we divide the test set into Head (top 20\%) and Tail (bottom 80\%). Note that the higher the performance on the tail items, the lower the popularity bias of the model.


\vspace{1mm}
\noindent
\textbf{Competing models}. 
To validate the effectiveness and efficiency of the proposed method, we adopt the following ten sequential recommendation (SR) models as baselines. The detailed descriptions of each model are in Section~\ref{sec:relatedwork}. 

\begin{itemize}[leftmargin=5mm]
    \item \textbf{Non-temporal SR models}: \textbf{SASRec}~\cite{KangM18SASRec} is a widely used baseline that utilizes a uni-directional self-attention mechanism, and \textbf{DuoRec}~\cite{QiuHYW22} improves SASRec through contrastive learning. \textbf{FEARec}~\cite{DuYZQZ0LS23} and \textbf{BSARec}~\cite{ShinCWP24} are state-of-the-art SR models that utilize frequency information to represent user sequences.

\vspace{0.5mm}
    \item \textbf{Temporal SR models}: \textbf{TiSASRec}~\cite{LiWM20} uses positional encoding with time interval information, and \textbf{TCPSRec}~\cite{TianLBWZ22} utilizes the invariance and periodicity of sequences to pre-train the model. \textbf{TiCoSeRec}~\cite{DangYGJ0XSL23} performs contrastive learning by augmenting uniform sequences with small time intervals.
    
\vspace{0.5mm}
    \item \textbf{Efficient SR models}: \textbf{LinRec}~\cite{LiuCZZGWLFWHLL23} proposes linear attention that can be applied to various transformer-based models, while \textbf{LRURec}~\cite{YueWHZMW24} converts RNNs into linear recurrent modules to improve their efficiency. \textbf{SLIST}\footnote{Interestingly, we found that SLIST and SLIT have the same performance on four datasets, \ie, ML-1M, Beauty, Toys, and Sports, indicating that SLIS is less helpful in SR. In Section~\ref{sec:results}, we will refer to it as SLIST, but it is SLIT for the four datasets.}~\cite{ChoiKLSL21} is a linear model that only considers sequential order but not temporal information.
\end{itemize}

\vspace{1mm}
\noindent
\textbf{Implementation details}. 
All experiments were performed using Recbole~\cite{ZhaoHPYZLZBTSCX22, XuTZZWZLTZHPZCW23} framework, which is an open-source library for recommendation systems\footnote{\url{https://github.com/RUCAIBox/RecBole}}. We set the maximum sequence length for neural models to 50, following the conventions~\cite{TianLBWZ22, DuYZQZ0LS23, DangYGJ0XSL23}. The other hyperparameters were tuned using the validation set, with an early stop based on NDCG@10. We performed a grid search with training batch size of \{128, 256, 512, 1024, 2048\}, learning rates of \{1e-4, 5e-4, 1e-3, 5e-3\}, number of layers of \{1, 2, 3\}, number of heads of \{1, 2, 3\}, and weight decay of \{1e-8, 0\}. The dropout rate was searched in the interval [0, 0.5] with step size 0.1, referring to~\cite{QiuHYW22}. For LinRec~\cite{LiuCZZGWLFWHLL23}, we use SASRec~\cite{KangM18SASRec} as a backbone. Each model's own hyperparameters were searched by following the original papers. The L2-regularization coefficient $\lambda$ of TALE was searched in \{1, 5, 10, 50, 100, 500, 1000\}. For the time interval-aware weighting, $\tau_{\text{time}}$ was searched in the range [$2^{-10}$, $2^{10}$] with exponentially increasing steps in powers of 2, and threshold $c$ was tuned in [0, 1] with step size 0.1. The window size $N$ (days) of trend-aware normalization was searched in \{7, 30, 90, 180, 360, 720\}.
We run all experiments on NVIDIA RTX-3090 24GB GPU and Intel Xeon Gold 6226R CPU.



\begin{table} 
\centering
\caption{Accuracy and efficiency comparison for efficient SR models on five datasets. `Train' and `Eval' mean the runtime (seconds) for training and evaluation. For linear models, the runtime is measured on CPU and for other baselines on GPU. }
\label{tab:efficient_models_comparison}
\vspace{-3mm}
\renewcommand{\arraystretch}{0.9} 

\begin{tabular}{c|c|cc|rr}
\toprule
Dataset & Model    & HR@5   & NDCG@5  & Train & Eval \\
\midrule
\multirow{4}{*}{ML-1M} 
& LinRec        & 0.1843    & 0.1254    & 1,241 & 10        \\
& LRURec        & 0.1491  & 0.0936  & 2,198     &   29        \\
& SLIST         & 0.1710  & 0.1164  & 412     &   \textbf{0.2}        \\
& TALE          & \textbf{0.1854}    & \textbf{0.1275}    & \textbf{179}  & \textbf{0.2}            \\
\midrule
\multirow{4}{*}{Beauty}
& LinRec        & 0.0550    & 0.0323    & 285 & 40            \\
& LRURec        & 0.0363  & 0.0257  & 678     &  233           \\
& SLIST         & 0.0597  & 0.0435  & \textbf{5}     &   \textbf{10}        \\
& TALE          & \textbf{0.0674}    & \textbf{0.0485}    & \textbf{5}  & \textbf{10}            \\
\midrule
\multirow{4}{*}{Toys}
& LinRec        &  0.0631  & 0.0359  &  197  &  37           \\
& LRURec        &  0.0624 &  0.0370  &    356   & 92           \\
& SLIST         & 0.0722  & 0.0535  & \textbf{5}     &   \textbf{9}        \\
& TALE          &  \textbf{0.0815} & \textbf{0.0594}  &  \textbf{5} &   \textbf{9}         \\
\midrule
\multirow{4}{*}{Sports}
& LinRec        &  0.0314  &  0.0172 & 529 &   52          \\
& LRURec        &  0.0178 &  0.0105  & 725 &  163          \\
& SLIST         & 0.0380  & 0.0271  & 14     &   \textbf{19}        \\
& TALE          & \textbf{0.0418} &  \textbf{0.0305}  &   \textbf{13} &  \textbf{19}          \\
\midrule
\multirow{4}{*}{Yelp}
& LinRec        & 0.0414    & 0.0316    & 230  & 41            \\
& LRURec        & 0.0269  & 0.0164  & 531       & 141            \\
& SLIST         & \textbf{0.0563}  & 0.0375  & 24     &   \textbf{21}        \\
& TALE          & 0.0558    & \textbf{0.0394}    & \textbf{13}  & \textbf{21}            \\
\bottomrule
\end{tabular}
\vspace{-2mm}

\end{table}

\section{Experimental Results}\label{sec:results}

\subsection{Performance Comparison}
\textbf{Comparison with traditional SR models}.
Table~\ref{tab:overall_perform} shows the performance comparison of TALE and the eight SR models, and we reveal three key findings. (i) Linear models (\ie, SLIST and TALE) achieve comparable to or better performance than neural SR models. In particular, they outperform the state-of-the-art SR model, BSARec, on four datasets (\ie, Beauty, Toys, Sports, and Yelp) which have short average sequence lengths and sparse interactions.
(ii) TALE consistently outperforms SLIST due to considering temporal information with an average gain of 9.07\% on NDCG@10. (Appendix~\ref{appen:reflect_tempo_info} shows that the learned weight matrix of TALE reflects more temporal information than that of SLIST.) Meanwhile, utilizing additional temporal information does not always result in higher performance. While TiSASRec and TCPSRec have higher performance than the backbone model (\ie, SASRec), in many cases, DuoRec, FEARec, and BSARec are superior to SR models with temporal information\footnote{For TiCoSeRec, despite the reference to the official source code (\url{https://github.com/KingGugu/TiCoSeRec}), we could not reproduce the performance on ML-1M and Yelp.}. Based on this result, we can see that temporal information is important in SR, as well as the model architecture and training loss. (iii) For ML-1M, neural models outperform linear models in some metrics. We assume this is because the neural models robustly train on highly dense ML-1M. Nevertheless, TALE performs better on some evaluation metrics (\eg, HR@1 and NDCG@5).

\vspace{1mm}
\noindent
\textbf{Comparison with efficient SR models}.
Table~\ref{tab:efficient_models_comparison} shows the accuracy and efficiency with efficient SR models (\ie, LinRec~\cite{LiuCZZGWLFWHLL23} and LRURec~\cite{YueWHZMW24}) on five datasets. For a fair comparison, we use batch sizes of 2048 and 1 for training and evaluation, respectively. For all five datasets, TALE performs the best compared to LinRec and LRURec, with the lowest training and evaluation time. Thanks to the closed-form solution, on ML-1M, Beauty, and Yelp, TALE achieves 6.9x, 57x, and 17.7x faster training time than LinRec, respectively. In a nutshell, TALE shows higher accuracy and superior efficiency compared to efficient neural SR models. Efficiency comparison with the other seven SR models is shown in Appendix~\ref{appen:efficiency_comparison}.

\begin{table} \small
\caption{Tail and Head performance comparison on ML-1M and Beauty. `Norm.' denotes the existing normalization method (\ie, Eq.~\eqref{eq:sym_norm}). Each metric is NDCG@5. }
\label{tab:tail_unbiased}
\vspace{-3mm}
\begin{center}
\renewcommand{\arraystretch}{1} 
\begin{tabular}{P{1.4cm}|P{0.75cm}P{0.75cm}P{0.75cm}|P{0.75cm}P{0.75cm}P{0.75cm}}
\toprule
Dataset          & \multicolumn{3}{c|}{ML-1M}          & \multicolumn{3}{c}{Beauty}               \\
Model            &    All      &   Tail      & Head   &    All   & Tail      & Head            \\
\midrule
SASRec           & 0.1197  & 0.0648    & 0.1567    & 0.0319          & 0.0167     & 0.0508        \\
BSARec           & \underline{0.1261}  & \underline{0.0740}    & \textbf{0.1642}  & 0.0338  & 0.0169    & 0.0548        \\
TiSASRec         & 0.1216  & 0.0706    & 0.1589     & 0.0340             & 0.0212       & 0.0499        \\
\midrule
SLIST            & 0.1164  & 0.0624    & 0.1559   & \underline{0.0435} & 0.0235    & \underline{0.0676}        \\
SLIST+Norm.      & 0.1030  & 0.0675    & 0.1290   & 0.0360 & \textbf{0.0275}    & 0.0463        \\
TALE             & \textbf{0.1275}  & \textbf{0.0814}    & \underline{0.1612}   & \textbf{0.0485} & \underline{0.0271}  & \textbf{0.0736}   \\

\bottomrule
\end{tabular}
\end{center}
\vspace{-2mm}
\end{table}



\begin{table} \small
\caption{Ablation study of TALE on ML-1M and Beauty. TALE consists of three parts: (1) single-target augmentation, (2) time interval-aware weighting, and (3) trend-aware normalization. Each metric is NDCG@5.}
\label{tab:ablation_performance}
\vspace{-3mm}
\begin{center}
\renewcommand{\arraystretch}{1} 
\begin{tabular}{P{0.25cm}P{0.25cm}P{0.25cm}|P{0.75cm}P{0.75cm}P{0.75cm}|P{0.75cm}P{0.75cm}P{0.75cm}}
\toprule
\multicolumn{3}{c|}{Dataset}         & \multicolumn{3}{c|}{ML-1M}          & \multicolumn{3}{c}{Beauty}              \\
(1) & (2) & (3)   & All   & Tail & Head  & All   & Tail & Head \\
\midrule
\cmark & \cmark     & \cmark  & \textbf{0.1275} & \textbf{0.0814} & \textbf{0.1614}   & \textbf{0.0485}   & \textbf{0.0271} & \underline{0.0736}     \\
- & \cmark     & \cmark   & 0.0473  & 0.0161 & 0.0701& 0.0296  & 0.0154 & 0.0467    \\
\cmark &     -      & \cmark  & 0.1227  & 0.0739 & 0.1584 & 0.0450  & 0.0239 & 0.0704     \\
\cmark & \cmark     & -  & \underline{0.1257}  & \underline{0.0773} & \underline{0.1611} & \underline{0.0468}  & \underline{0.0243} & \textbf{0.0739}       \\
-      &  -         & \cmark   & 0.1187  & 0.0659 & 0.1573 & 0.0438  & 0.0242  & 0.0663 \\
-      & \cmark     & -   & 0.0430  & 0.0114 & 0.0661 & 0.0303  & 0.0150  & 0.0487 \\
\cmark &     -      & -  & 0.1172  & 0.0685 & 0.1528 & 0.0444 & 0.0235  & 0.0696    \\
-      & -          & -   & 0.1164  & 0.0624 & 0.1559 & 0.0435  & 0.0235  & 0.0676 \\

\bottomrule
\end{tabular}
\vspace{-2mm}
\end{center}
\end{table}



\subsection{In-depth Analysis}\label{sec:exp_analysis}

In this section, we provide the empirical results on ML-1M and Beauty due to space limitations. The results on additional datasets (\ie, Toys, Sports, and Yelp) are shown in Appendix~\ref{appen:tail_head_perfrom}--\ref{appen:hyper_sensi}.

\vspace{1mm}
\noindent
\textbf{Popularity bias}.
To validate the ability of TALE to mitigate popularity bias, we split the entire test set into Head and Tail, and Table~\ref{tab:tail_unbiased} shows the performance for each. (i) TALE performs better than others on both tail and head items. In particular, TALE shows more performance gains for tail items, indicating that it is better at reducing popularity bias than neural models. Specifically, on ML-1M and Beauty, TALE achieves 30.4\% and 15.3\% improvement gains in a tail item performance over SLIST, respectively. (We will explore it in more detail in the ablation study.) (ii) Applying traditional normalization (Eq.~\eqref{eq:sym_norm}) to SLIST (\ie, SLIST+Norm.) improves performance for tail items, but it severely degrades performance for head items, resulting in significantly lower overall performance. This is because Eq.~\eqref{eq:sym_norm} only considers the cumulative popularity over the entire period, not the trend-aware popularity, which simply leads to recommendations towards unpopular items.

\vspace{1mm}
\noindent
\textbf{Ablation study}.
Table~\ref{tab:ablation_performance} indicates a breakdown of the effectiveness of each component of TALE. The key observations are as follows.
(i) Using all three components is more accurate than using only some of them. Using (1) and (2) together yields the largest performance gains among the partial combinations. Furthermore, using (1) leads to consistent performance improvements in all cases, as it is effective in learning the item relationship because it lowers the weight of items that are distant.
(ii) On the other hand, (2) without using (1) results in significant performance degradation. As shown in the theoretical analysis in Section~\ref{sec:single_aug}, not using (1), \ie, using the multi-target augmentation, cannot correctly learn the relationship between items.
(iii) (3) shows the debiasing effect, with tail item performance gains of 5.61\% and 2.98\% for ML-1M and Beauty, respectively, compared to not using (3). For datasets sensitive to item trends (\eg, ML-1M), applying (3) improves a head item performance besides tail items. It not only removes popularity bias in trend-sensitive datasets but also detects head items that users don't prefer. It indicates that (3) is possible to satisfy both trend-sensitive and non-trend-sensitive users.

\begin{figure}
\begin{tabular}{cc}
\includegraphics[height=2.75cm]{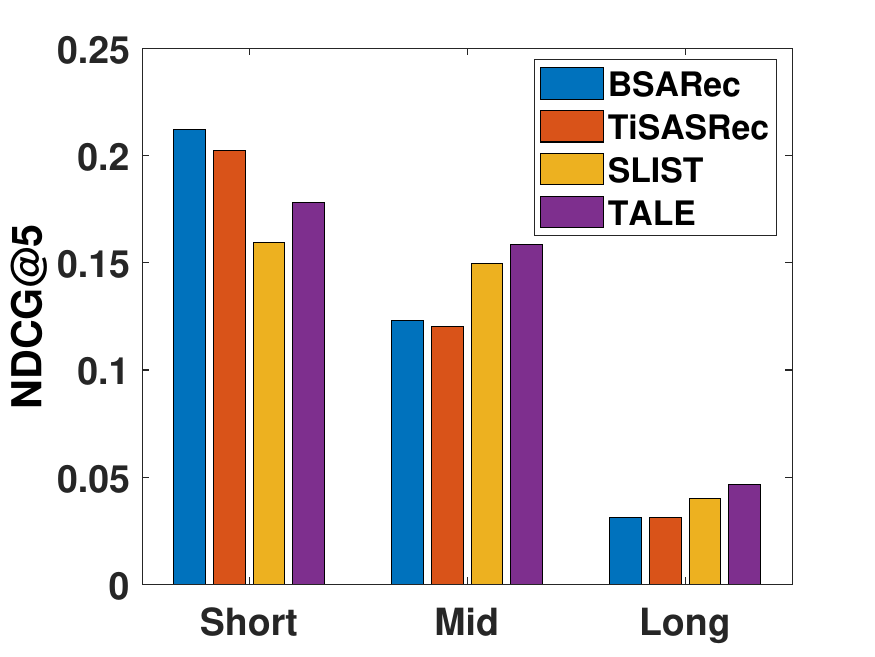} &
\includegraphics[height=2.75cm]{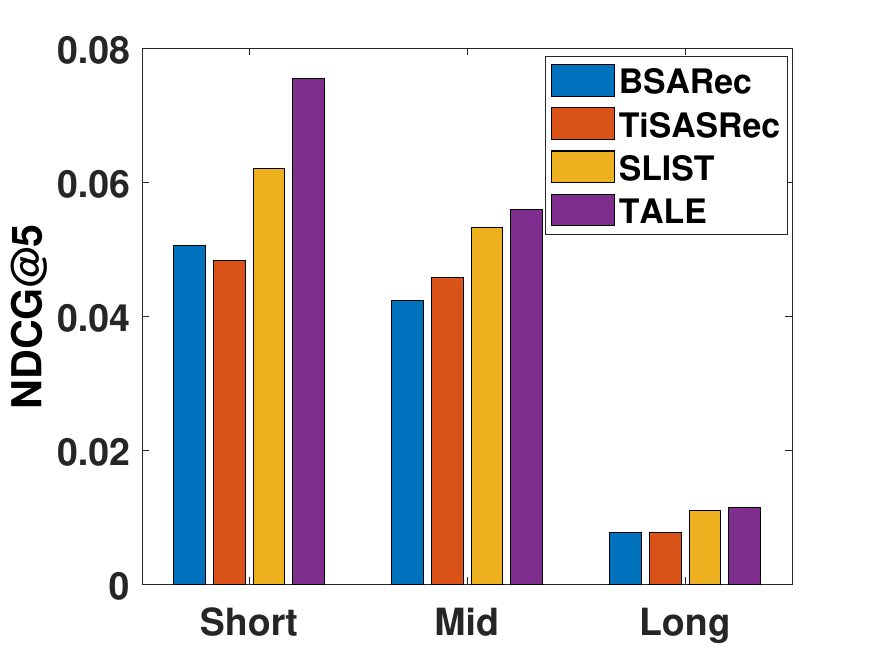} \\
\multicolumn{1}{c}{(a) ML-1M} &\multicolumn{1}{c}{(b) Beauty} \\
\end{tabular}
\vspace{-3mm}
\caption{Accuracy comparison by the time interval group on ML-1M and Beauty. }\label{fig:ablation_time_intv_perf}
\vspace{-5mm}
\end{figure}

\vspace{1mm}
\noindent
\textbf{Performance over time intervals}.
Figure~\ref{fig:ablation_time_intv_perf} shows performance based on the various time intervals between the test item and the validation item for each user. 
We divided the test set into three groups using time intervals: Short (0\% to 33.3\%), Mid (33.3\% to 66.6\%), and Long (66.6\% to 100\%). (i) All SR models tend to perform worse as the time interval increases. As the time interval increases, the user interest varies widely, which increases the difficulty of preference prediction. (ii) Nevertheless, TALE outperforms the baselines at relatively long time intervals (\ie, Mid and Long), showing that it effectively utilizes the long-time dependency. (iii) TALE outperforms SLIST in all subsets, which is due to its better capture of item relevance based on the use of temporal information.

\subsection{Hyperparameter Sensitivity} \label{sec:hyper_sensi}
\begin{figure}
\begin{tabular}{cc}
\includegraphics[height=3.0cm]{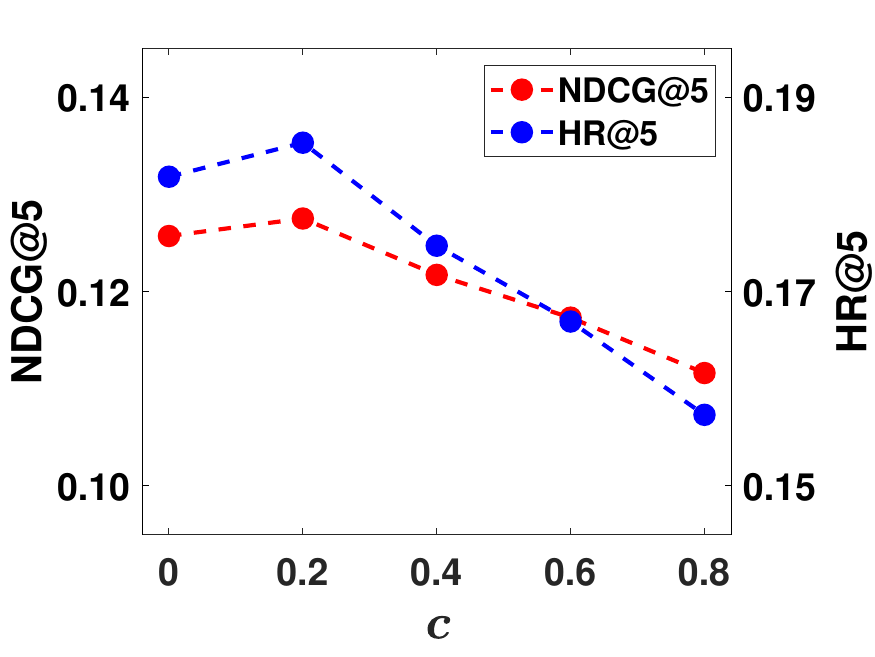} &
\includegraphics[height=3.0cm]{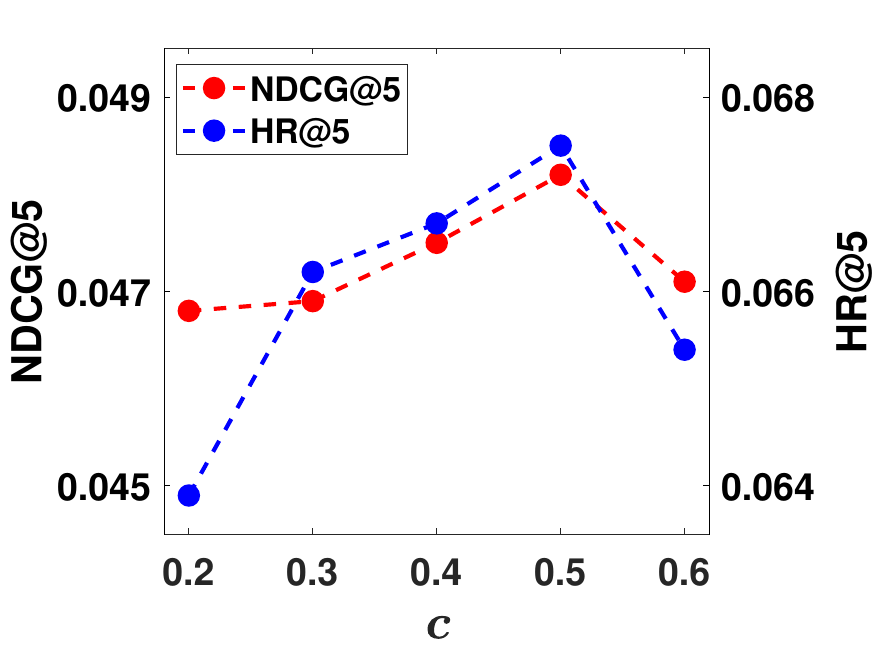} \\
\includegraphics[height=3.0cm]{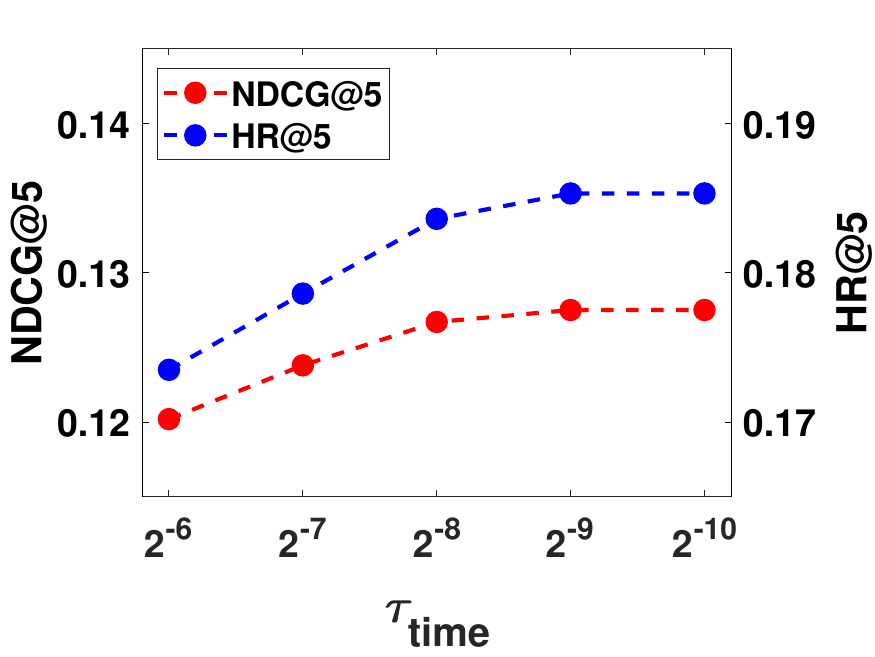} &
\includegraphics[height=3.0cm]{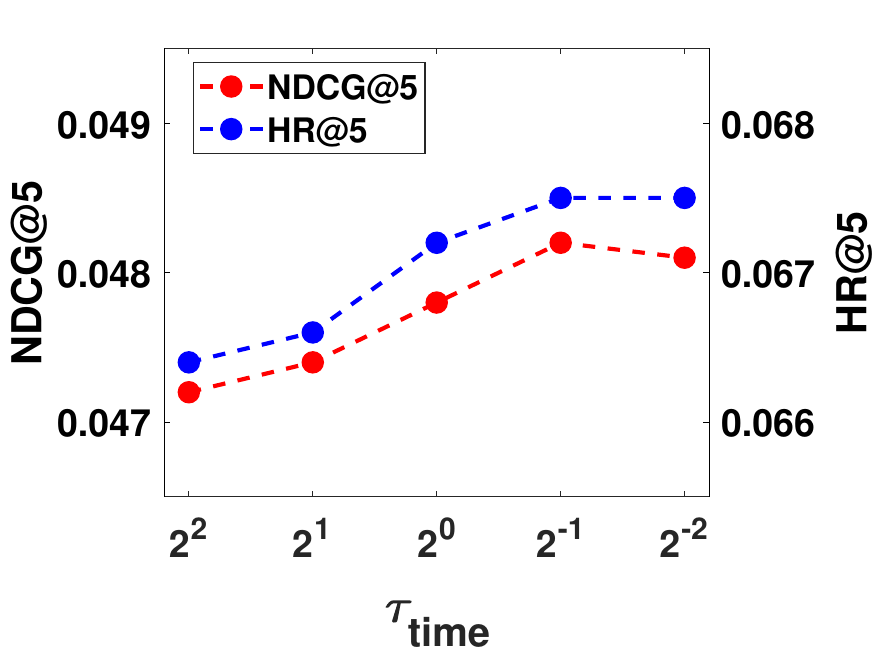} \\
\includegraphics[height=3.0cm]{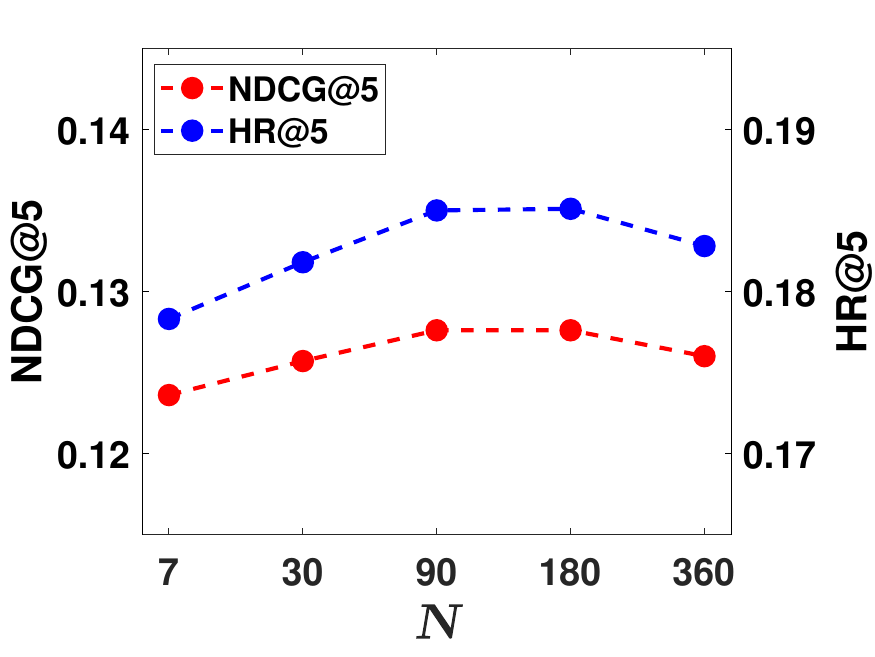} &
\includegraphics[height=3.0cm]{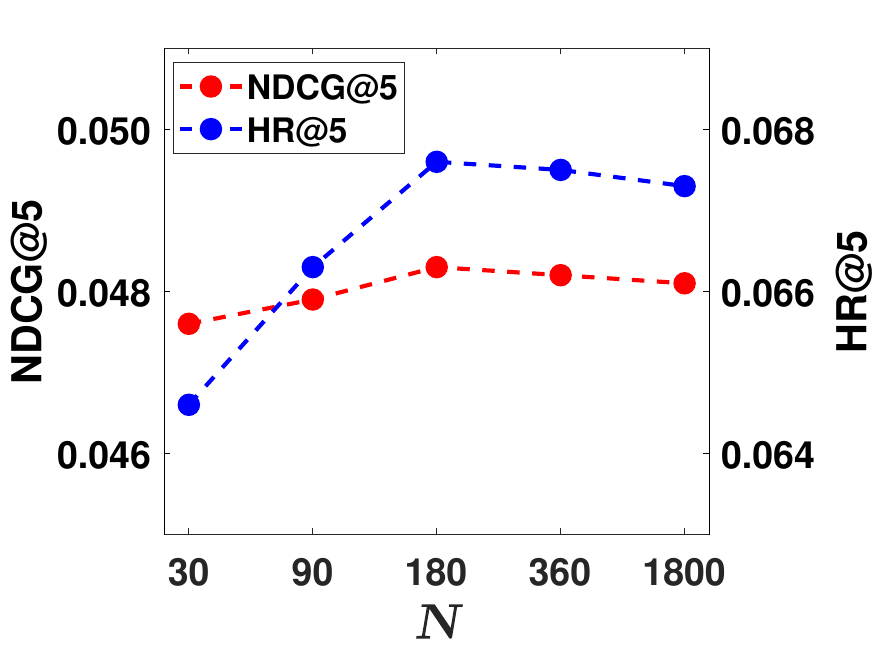} \\
\multicolumn{1}{c}{(a) ML-1M} &\multicolumn{1}{c}{(b) Beauty} \\
\end{tabular}
\vspace{-3mm}
\caption{Performance of TALE over $c$, $\tau_\text{time}$, and $N$ on ML-1M and Beauty.}\label{fig:hyper_c}

\vspace{-2mm}
\end{figure}

\textbf{Time interval-aware weighting}.
Figure~\ref{fig:hyper_c} provides the performance of TALE depending on threshold $c$ and time decay $\tau_{\text{time}}$, respectively. Both hyperparameters have different optimal values depending on the dataset. ML-1M has a smaller $\tau_{\text{time}}$ than Beauty because the average time interval in ML-1M is relatively small, about 14 hours. Also, threshold $c$ is greater in Beauty than in ML-1M, where long-time dependency is important. As such, time interval-aware weighting has to be adaptively applied to the characteristics of the dataset by tuning $c$ and $\tau_{\text{time}}$.

\vspace{1mm}
\noindent
\textbf{Trend-aware normalization}.
Figure~\ref{fig:hyper_c} shows the performance of TALE for different window sizes $N$. As the window size $N \rightarrow \infty$, it is equivalent to general normalization (Eq.~\eqref{eq:general_popularity}), so we can see that the performance converges as $N$ gets larger. We can observe that the optimal window size depends on the dataset; for example, the optimal $N$ is relatively small for ML-1M, which has a small average time interval. The empirical results of the three hyperparameters on Toys, Sports, and Yelp are in Appendix~\ref{appen:hyper_sensi}.

\section{Related Work} \label{sec:relatedwork}
We review sequential recommendation (SR) models in two groups: linear and neural SR models, as summarized in Table~\ref{tab:main_differences}.

\vspace{1mm}
\noindent
\textbf{Linear SR models}. They are classified into Markov chains (MC), neighborhood-based, and item-to-item models. FPMC~\cite{RendleFS10} combines MC with matrix factorization for capturing sequential dependency and user-item similarity. Sequential rules~\cite{KamehkhoshJL17} integrate MC and association rules~\cite{AgrawalIS93AR}. MC-based models are limited to capturing only simple patterns because they do not account for long-term dependencies between items. SKNN~\cite{JannachL17} introduces a K-nearest neighbor (KNN) approach to perform SR, utilizing K user histories that are similar to the target user. STAN~\cite{GargGMVS19} introduces weighting methods to SKNN~\cite{JannachL17} that reflects the sequential and temporal information. The recently proposed SLIST~\cite{ChoiKLSL21} is a linear item-to-item model with closed-form solutions that learn item similarity and transition information. S-Walk~\cite{ChoiKLSL22SWalk} is an improved version of SLIST~\cite{ChoiKLSL21}, which performs random walks with restarts by alternating between transition graphs and teleportation graphs. Although these linear SR models are more efficient than neural SR models, they do not yet take temporal information into account.


\begin{table}
\caption{Categorization of representative sequential recommender models and our proposed model.}\label{tab:main_differences}
\vspace{-3mm}
\renewcommand{\arraystretch}{1} 
\begin{tabular}{c|c|c}
\toprule
Type & Neural model & Linear model \\
\midrule
\begin{tabular}[c]{@{}c@{}} Sequential \\ info. \end{tabular} & \begin{tabular}[c]{@{}c@{}} GRU4Rec~\cite{HidasiKBT15GRU4Rec}, BERT4Rec~\cite{SunLWPLOJ19BERT4Rec}, \\ SASRec~\cite{KangM18SASRec}, DuoRec~\cite{QiuHYW22}, \\ LinRec~\cite{LiuCZZGWLFWHLL23}, FEARec~\cite{DuYZQZ0LS23}, \\ LRURec~\cite{YueWHZMW24}, BSARec~\cite{ShinCWP24} \end{tabular} & \begin{tabular}[c]{@{}c@{}} SKNN~\cite{JannachL17}, \\ STAN~\cite{GargGMVS19}, \\ SLIST~\cite{ChoiKLSL21} \end{tabular}\\ 
\midrule
\begin{tabular}[c]{@{}c@{}} Temporal \\ info. \end{tabular} & \begin{tabular}[c]{@{}c@{}} TiSASRec~\cite{LiWM20}, TCPSRec~\cite{TianLBWZ22}, \\ TGSRec~\cite{FanLZX0Y21}, TiCoSeRec~\cite{DangYGJ0XSL23}, \\
TGCL4SR~\cite{ZhangCWLX24} \end{tabular} & \textbf{TALE (Ours)} \\
\bottomrule
\end{tabular}
\vspace{-2mm}
\end{table}

\vspace{1mm}
\noindent
\textbf{Neural SR models}.
They can be categorized by whether they use temporal information. First, SR models without temporal information leverage sequential order among items in user history. GRU4Rec~\cite{HidasiKBT15GRU4Rec} utilizes gated recurrent units (GRU) to effectively capture short and long-term dependencies between items. SASRec~\cite{KangM18SASRec} introduces transformers~\cite{VaswaniSPUJGKP17} to SR, followed by various neural SR models. BERT4Rec~\cite{SunLWPLOJ19BERT4Rec} improves on SASRec by leveraging bi-directional self-attention, and DuoRec~\cite{QiuHYW22}, CoSeRec~\cite{abs-2108-06479}, and CL4SRec~\cite{XieSLWGZDC22} further utilize multi-task learning with contrastive loss for better sequence representation. More recently, FMLPRec~\cite{ZhouYZW22}, FEARec~\cite{DuYZQZ0LS23}, and BSARec~\cite{ShinCWP24} interpret SR from a frequency domain perspective to capture user preferences that are difficult to capture with self-attention alone. However, these models that do not utilize temporal information suffer because they cannot fully capture the item relationships in user history.


Temporal neural SR models show enhanced performance by leveraging absolute timestamps to recommend timely items. TiSASRec~\cite{LiWM20} improves SASRec~\cite{KangM18SASRec} by introducing relative time interval embeddings and utilizing them as the key and value for self-attention. TGSRec~\cite{FanLZX0Y21} proposes the temporal collaborative transformer using additional temporal embeddings with absolute time. TCPSRec~\cite{TianLBWZ22} models the invariant and periodic properties of sequences and performs pre-training by contrastive learning using them. TiCoSeRec~\cite{DangYGJ0XSL23} improves CoSeRec~\cite{abs-2108-06479} by increasing the quality of augmented sequences by ensuring that the standard deviation of the time interval is smaller. Finally, TGCL4SR~\cite{ZhangCWLX24} performs two graph augmentations based on the user-item bipartite graph with timestamps to generate various subgraphs.

Most neural SR models show high performance by exploiting complex patterns between items and sequences but are limited by low efficiency. To address this limitation, recent studies~\cite{LiuCZZGWLFWHLL23, YueWHZMW24} have attempted to reduce the computation of self-attention. LinRec~\cite{LiuCZZGWLFWHLL23} proposes linear attention that can be applied to various transformers-based models, and LRURec~\cite{YueWHZMW24} converts RNNs into linear recurrent modules to improve their efficiency. However, they cannot dramatically reduce the amount of computation due to the complex nature of deep neural networks.

\section{Conclusion}\label{sec:conclusion}

This paper proposed an efficient linear SR model \textbf{TALE}, which successfully utilizes temporal context via three key components: \emph{single-target augmentation}, \emph{time interval-aware weighting}, and \emph{trend-aware normalization}. Across five datasets, TALE consistently demonstrated superior accuracy compared to ten other SR models, improving HR@1 by up to 25\%. Owing to trend-aware normalization, TALE showed outstanding accuracy for both head and tail items. All components in TALE can be pre-computed, so no additional training and inference time was required. Notably, it achieved 17.7x faster training time and 145x faster inference time than existing efficient neural SR models.

In future work, we can enhance each proposed method as follows. (i) We will further employ a range of temporal data augmentations~\cite{DangYGJ0XSL23, QiuHYW22, abs-2108-06479} for constructing training matrices for linear SR models. (ii) We will identify various consumption patterns of items through statistical analysis of the datasets and design new popularity functions guided by these insights to perform normalization.

\begin{acks}
    This work was supported by Institute of Information \& communications Technology Planning \& evaluation (IITP) grant and National Research Foundation of Korea (NRF) grant funded by the Korea government (MSIT) (No. RS-2019-II190421, RS-2020-II201821, and RS-2021-II212068).
\end{acks}

\balance

\appendix

\section{Mathematical Proof} \label{appen:math_proof}

\subsection{Multi-Target Augmentation} \label{appen:proof_multi_target}

Through the following expansion, we prove that multi-target augmentation learns more items with large position gaps.
\begin{align} 
    & \hat{\mathbf{B}}_{\text{multi}} = \underset{\mathbf{B}}{\text{argmin}}\sum_{u'=1}^{m'} \|\mathbf{T}_{u',*}^{\text{multi}}-\mathbf{S}_{u',*}\mathbf{B}\|^2_F \\
     & = \underset{\mathbf{B}}{\text{argmin}}\sum_{u=1}^{m}\sum_{l=1}^{|\mathcal{S}^u|-1}\| \mathbf{x}(i_{l+1:|\mathcal{S}^u|}^u) - \mathbf{x}(i_{1:l}^u)\mathbf{B}  \|^2_{F} \label{eq:slit_objective_2} \\
    & = \underset{\mathbf{B}}{\text{argmin}}\sum_{u=1}^{m}\sum_{l=1}^{|\mathcal{S}^u|-1}\Big\{\sum_{s=1}^{l}\Big(\| \mathbf{x}(i_{l+1:|\mathcal{S}^u|}^{u}) - \mathbf{x}(i_{s}^{u})\mathbf{B} \|^2_{F} \nonumber \\
    & - \|\mathbf{x}(i_{s}^u)\mathbf{B}  \|^2_{F}\Big) + \| \mathbf{x}(i_{1:l}^{u})\mathbf{B}\|^2_{F} \Big\} \label{eq:slit_objective_3} \\
    & =  \underset{\mathbf{B}}{\text{argmin}}\sum_{u=1}^{m}\sum_{l=1}^{|\mathcal{S}^u|-1}\sum_{s=1}^{l}\| \mathbf{x}(i_{l+1:|\mathcal{S}^u|}^{u}) - \mathbf{x}(i_{s}^{u})\mathbf{B}\|^2_{F} + \Tilde{\textbf{B}}_{1}^{S} \label{eq:slit_objective_4}\\
    & = \underset{\mathbf{B}}{\text{argmin}}\sum_{u=1}^{m}\sum_{l=1}^{|\mathcal{S}^u|-1}\Big\{\sum_{s=1}^{l}\Big(\sum_{h=l+1}^{|\mathcal{S}^u|}\|\mathbf{x}(i_{h}^u) - \mathbf{x}(i_{s}^u)\mathbf{B}\|^2_{F}\Big) \nonumber \\
    & - (|\mathcal{S}^u|-l-1)\|\mathbf{x}(i_{s}^u)\mathbf{B}\|^2_{F}\Big\} + \Tilde{\textbf{B}}_{1}^{S} \label{eq:slit_objective_5} \\
    & = \underset{\mathbf{B}}{\text{argmin}}\sum_{u=1}^{m}\sum_{l=1}^{|\mathcal{S}^u|-1}\sum_{s=1}^{l}\sum_{h=l+1}^{|\mathcal{S}^u|}\|\mathbf{x}(i_{h}^u) - \mathbf{x}(i_{s}^u)\mathbf{B}\|^2_{F} + \Tilde{\textbf{B}}_{2}^{S} \label{eq:slit_objective_6}\\
    & = \underset{\mathbf{B}}{\text{argmin}}\sum_{u=1}^{m}\sum_{s=1}^{|\mathcal{S}^u|-1}\sum_{h=s+1}^{|\mathcal{S}^u|} (h-s)\|\mathbf{x}(i_{h}^u) - \mathbf{x}(i_{s}^u)\mathbf{B}\|^2_{F} + \Tilde{\textbf{B}}_{2}^{S},\label{eq:slit_objective_7}
\end{align}

\begin{equation}
\begin{aligned}
    \text{where}~\Tilde{\textbf{B}}_{1}^{S} = \sum_{u=1}^{m}\sum_{l=1}^{|\mathcal{S}^u|-1} \|\mathbf{x}(i_{1:l})\mathbf{B}\|^2_F - \sum_{u=1}^{m}\sum_{l=1}^{|\mathcal{S}^u|-1}\sum_{s=1}^{l}\|\mathbf{x}(i_{s}^{u})\mathbf{B}\|^2_F, \\
    \Tilde{\textbf{B}}_{2}^{S} = \sum_{u=1}^{m}\sum_{l=1}^{|\mathcal{S}^u|-1} \|\mathbf{x}(i_{1:l})\mathbf{B}\|^2_F - \sum_{u=1}^{m}\sum_{l=1}^{|\mathcal{S}^u|-1}\sum_{s=1}^{l}(|\mathcal{S}^u| - l)\|\mathbf{x}(i_{s}^{u})\mathbf{B}\|^2_F. \nonumber
\end{aligned}
\end{equation}

Eq.~\eqref{eq:slit_objective_2} is the objective equation for the optimization problem with multi-target augmentation. For a user $u$, given items up to $l$-th in the sequence, it learns to restore items after ($l+1$)-th.
To further analyze the learning between the source and target item, we first convert Eq.~\eqref{eq:slit_objective_2} into Eq.~\eqref{eq:slit_objective_3} which has only one source item. Then, we transform the equation using the following process.

\begin{itemize}[leftmargin=5mm]
    \item \texttt{\textbf{Step1}}: $\text{argmin}_{\mathbf{D}}\|\mathbf{C}-(\mathbf{z}_1+\mathbf{z}_2)\mathbf{D}\|^2_F=\text{argmin}_{\mathbf{D}}\|\mathbf{C}\|^2_F + \|(\mathbf{z}_1+\mathbf{z}_2)\mathbf{D}\|^2_F - \mathbf{C}^{\top}(\mathbf{z}_1+\mathbf{z}_2)\mathbf{D} - ((\mathbf{z}_1+\mathbf{z}_2)\mathbf{D})^{\top}\mathbf{C}$.
\vspace{0.5mm}
    \item \texttt{\textbf{Step2}}: $\text{argmin}_{\mathbf{D}}\|\mathbf{C}\|^2_F + \|(\mathbf{z}_1+\mathbf{z}_2)\mathbf{D}\|^2_F - \mathbf{C}^{\top}(\mathbf{z}_1+\mathbf{z}_2)\mathbf{D} - ((\mathbf{z}_1+\mathbf{z}_2)\mathbf{D})^{\top}\mathbf{C} = \text{argmin}_{\mathbf{D}}\|\mathbf{C}\|^2_F + \|(\mathbf{z}_1+\mathbf{z}_2)\mathbf{D}\|^2_F - \mathbf{C}^{\top}(\mathbf{z}_1+\mathbf{z}_2)\mathbf{D} - ((\mathbf{z}_1+\mathbf{z}_2)\mathbf{D})^{\top}\mathbf{C} + \|\mathbf{C}\|^2_F$.
\vspace{0.5mm}
    \item \texttt{\textbf{Step3}}: $\text{argmin}_{\mathbf{D}}2\|\mathbf{C}\|^2_F + \|(\mathbf{z}_1+\mathbf{z}_2)\mathbf{D}\|^2_F - \mathbf{C}^{\top}(\mathbf{z}_1+\mathbf{z}_2)\mathbf{D} - ((\mathbf{z}_1+\mathbf{z}_2)\mathbf{D})^{\top}\mathbf{C} = \text{argmin}_{\mathbf{D}}\big(\|\mathbf{C}\|^2_F - \mathbf{C}^{\top}\mathbf{z}_1\mathbf{D} - \mathbf{D}^{\top}\mathbf{z}_1^{\top}\mathbf{C} +\|\mathbf{z}_1\mathbf{D}\|^2_F\big) + \big(\|\mathbf{C}\|^2_F - \mathbf{C}^{\top}\mathbf{z}_2\mathbf{D} - \mathbf{D}^{\top}\mathbf{z}_2^{\top}\mathbf{C} +\|\mathbf{z}_2\mathbf{D}\|^2_F\big) + \big(\|(\mathbf{z}_1+\mathbf{z}_2)\mathbf{D}\|^2_F - \|\mathbf{z}_1\mathbf{D}\|^2_F-\|\mathbf{z}_2\mathbf{D}\|^2_F\big) = \text{argmin}_{\mathbf{D}}\|\mathbf{C}-\mathbf{z}_1\mathbf{D}\|^2_F + \|\mathbf{C}-\mathbf{z}_2\mathbf{D}\|^2_F + \|(\mathbf{z}_1+\mathbf{z}_2)\mathbf{D}\|^2_F - \|\mathbf{z}_1\mathbf{D}\|^2_F-\|\mathbf{z}_2\mathbf{D}\|^2_F$.
\end{itemize}

Finally, we derived Eq.~\eqref{eq:slit_objective_4} to consider only one source item.
From this point on, the terms that are only affected by the source items are denoted by $\Tilde{\mathbf{B}}^{S}_{*}$.\footnote{Single- and multi-target augmentation have the same estimate of the concentration matrix $(\mathbf{S}^{\top}\mathbf{S})^{-1}$.} Then, we transform Eq.~\eqref{eq:slit_objective_4} into Eq.~\eqref{eq:slit_objective_5} which has only one target item.\footnote{It is derived through a similar process of converting Eq.~\eqref{eq:slit_objective_2} to Eq.~\eqref{eq:slit_objective_3}.}  In Eq.~\eqref{eq:slit_objective_6}, $l$ determines the range of $s$ and $h$, and also determines the number of times the relationship between the source item and the target item is learned. To be more specific, the relationship between source item $i_1^u$ and target item $i_{|\mathcal{S}^u|}^u$ is learned when $l$ has a value between 1 and $|\mathcal{S}^u|$, and the relationship between source item $i_2^u$ and target item $i_{|\mathcal{S}^u|-1}^u$ is learned when $l$ has a value between 2 and $|\mathcal{S}^u|-1$. This pattern helps us to know how many times the relationship between source item $i_h^u$ and target item $i_s^u$ is learned. So, we can convert Eq.~\eqref{eq:slit_objective_6} to Eq.~\eqref{eq:slit_objective_7} using this pattern.

In conclusion, the objective function of multi-target augmentation assigns weight according to the position gap (\ie, $h-s$) in learning the relationship between the target and source item. This is a problem when applying temporal weights. In particular, weighting by position gap can lead linear model learning to be biased towards long sequences. Consequently, we need to design a new augmentation method that is well-suited to learning temporal information.

\subsection{Single-Target Augmentation} \label{appen:proof_single_target}
The derivation of single-target augmentation is similar to that of multi-target augmentation. However, unlike multi-target augmentation, it is not weighted by position gap, meaning that it is a more suitable approach to inject temporal information into linear models.
\begin{align} 
    \hat{\mathbf{B}}_{\text{single}} 
    & = \underset{\mathbf{B}}{\text{argmin}}\sum_{u'=1}^{m'} \|\mathbf{T}_{u',*}^{\text{single}}-\mathbf{S}_{u',*}\mathbf{B}\|^2_F \\
    & = \underset{\mathbf{B}}{\text{argmin}}\sum_{u=1}^{m}\sum_{h=1}^{|\mathcal{S}^u|-1}\|\mathbf{x}(i_{l+1}^u)-\mathbf{B}  \mathbf{x}(i_{1:l}^u)\|^2_{F} \label{eq:tale_objective_2} \\
    & = \underset{\mathbf{B}}{\text{argmin}}\sum_{u=1}^{m}\sum_{l=1}^{|\mathcal{S}^u|-1}\Big\{\sum_{s=1}^{l}\Big(\| \mathbf{x}(i_{l+1}^{u}) - \mathbf{x}(i_{s}^{u})\mathbf{B} \|^2_{F} \nonumber \\
    & - \|\mathbf{x}(i_{s}^u)\mathbf{B}  \|^2_{F}\Big) + \| \mathbf{x}(i_{1:l}^{u})\mathbf{B}\|^2_{F} \Big\} \label{eq:tale_objective_3}\\
    & =  \underset{\mathbf{B}}{\text{argmin}}\sum_{u=1}^{m}\sum_{l=1}^{|\mathcal{S}^u|-1}\sum_{s=1}^{l}\| \mathbf{x}(i_{l+1}^{u}) - \mathbf{x}(i_{s}^{u})\mathbf{B}\|^2_{F} + \Tilde{\textbf{B}}_{1}^{S} \label{eq:tale_objective_4}\\
    & = \underset{\mathbf{B}}{\text{argmin}}\sum_{u=1}^{m}\sum_{s=1}^{|\mathcal{S}^u|-1}\sum_{h=s+1}^{|\mathcal{S}^u|} \|\mathbf{x}(i_{h}^u) - \mathbf{x}(i_{s}^u)\mathbf{B}\|^2_{F} + \Tilde{\textbf{B}}_{1}^{S}. \label{eq:tale_objective_5}
\end{align}

\begin{table}[t]
\caption{The best hyperparameters of TALE on five datasets.}
\label{tab:best_hyper}
\vspace{-2mm}
\begin{center}
\begin{tabular}{c|ccccc}
\toprule
Dataset          & $\lambda$       & $c$ & $\tau_{\text{time}}$ & $N$ \\
\midrule 
ML-1M            & 100             & 0.2      & 1/512  & 180    \\ 
Beauty           & 100             & 0.4      & 1/2    & 180    \\ 
Toys             & 100             & 0.3      & 1      & 360    \\ 
Sports           & 500             & 0.4      & 4      & 180    \\ 
Yelp             & 0.001           & 0.2      & 1/32   & 180    \\ 
\bottomrule
\end{tabular}
\end{center}
\end{table}

\section{Detailed Experimental Setup} \label{appen:detailed_exp_setup}

\subsection{Hyperparameter Settings} \label{appen:best_hyperparameters}
Table~\ref{tab:best_hyper} suggests the best hyperparameters of TALE. For Yelp, we only utilize a combination of TALE and SLIS and set $\alpha$ to 0.9. In~\cite{ChoiKLSL21}, $\alpha \in [0,1]$ is used to control the importance between SLIS and SLIT; if $\alpha$ is set to zero, only SLIT is used.

\section{Additional Experimental Results} \label{appen:add_exp}

This section contains the following experiments.
\begin{itemize}[leftmargin=5mm]
    \item Appendix~\ref{appen:transition_prob} verifies the existence of user preference drifts.
    
    \vspace{0.5mm}
    \item Appendix~\ref{appen:effect_trend_norm} shows the debiasing effect of TALE.

    \vspace{0.5mm}
    \item Appendix~\ref{appen:reflect_tempo_info} demonstrates how well the learned item-to-item weight matrix $\hat{\mathbf{B}}$ incorporates temporal information (\ie, user preference drifts over time).

    \vspace{0.5mm}
    \item Appendix~\ref{appen:efficiency_comparison} shows that the efficiency comparison between TALE and seven SR models except for efficient SR models (\ie, LinRec~\cite{LiuCZZGWLFWHLL23}, LRURec~\cite{YueWHZMW24}, and SLIST~\cite{ChoiKLSL21}).

    \vspace{0.5mm}
    \item Appendix~\ref{appen:intv_co_occur},~\ref{appen:tail_head_perfrom}--\ref{appen:hyper_sensi} present the experimental results on the remaining three datasets (\ie, Toys, Sports, and Yelp) that we were unable to include in the main text due to space limits.
    
\end{itemize}


\begin{figure}
\begin{tabular}{cc}
\includegraphics[height=3.0cm]{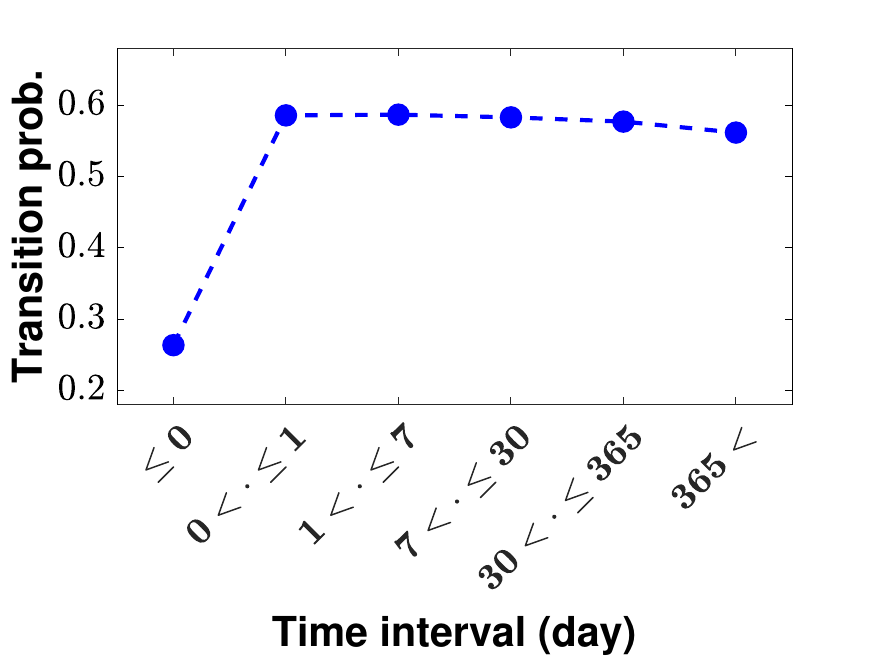} &
\includegraphics[height=3.0cm]{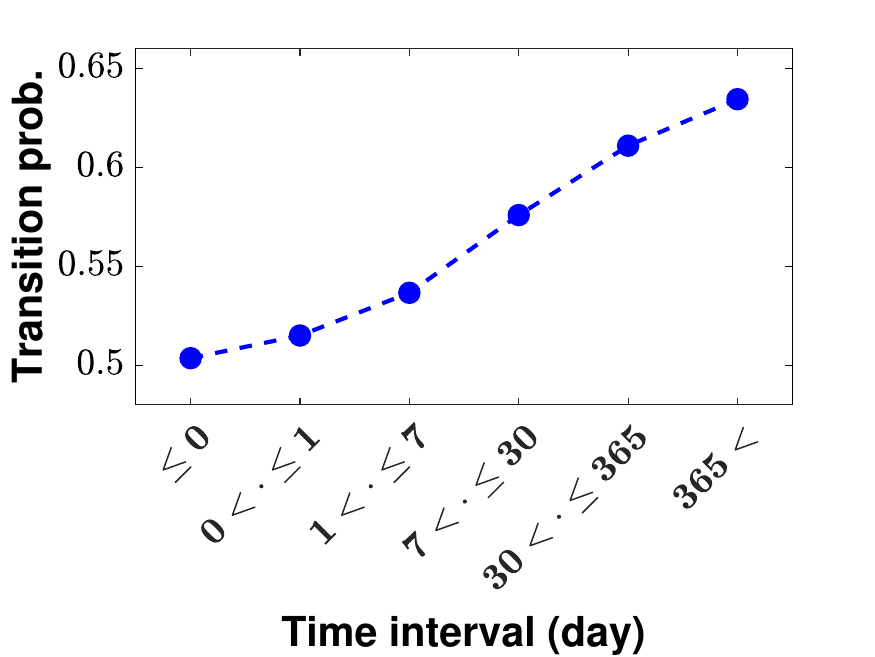} \\
\multicolumn{1}{c}{(a) ML-1M} &\multicolumn{1}{c}{(b) Beauty} \\
\end{tabular}

\vspace{2mm}

\begin{tabular}{cc}
\includegraphics[height=3.0cm]{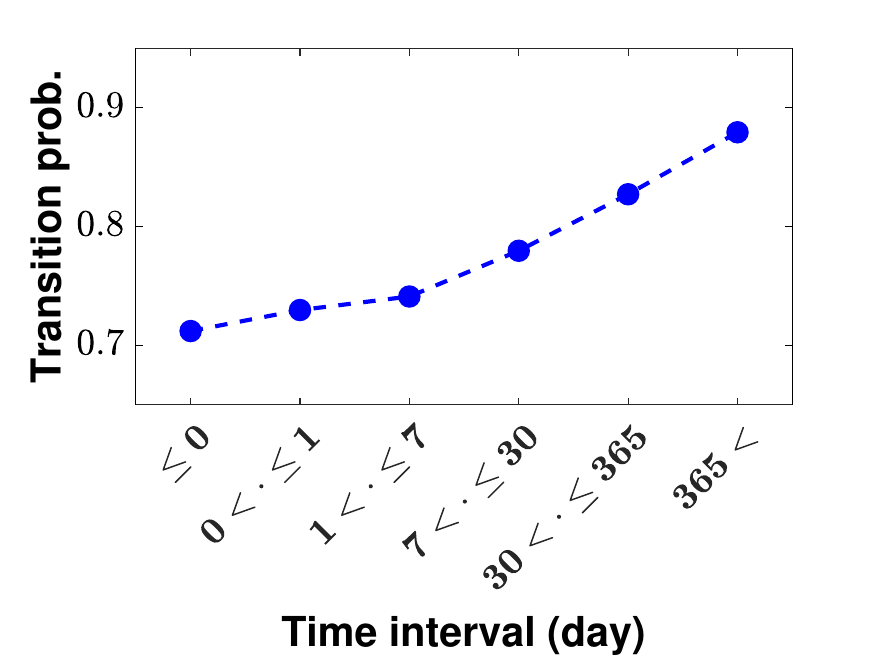} &
\includegraphics[height=3.0cm]{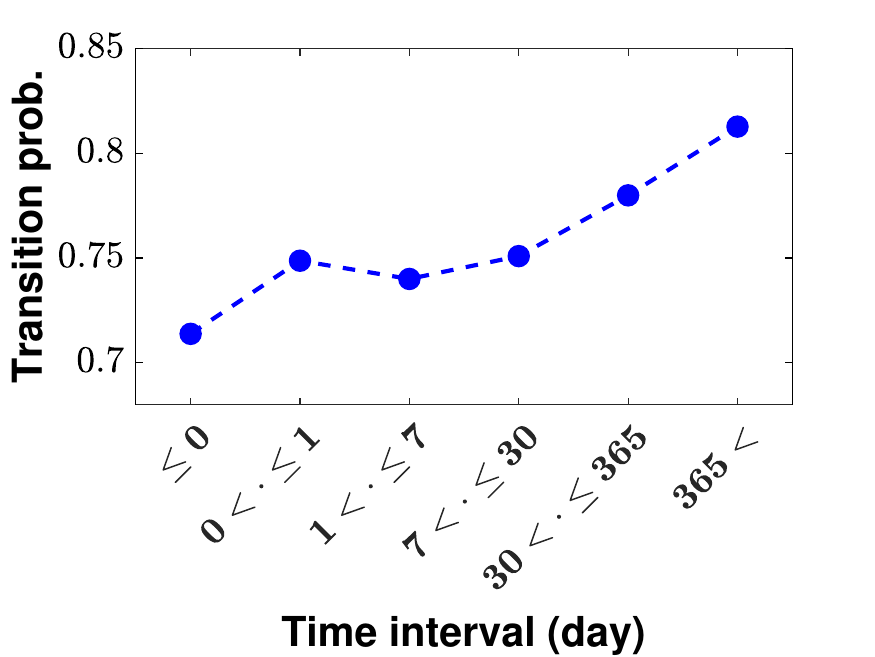} \\
\multicolumn{1}{c}{(c) Toys} &\multicolumn{1}{c}{(d) Sports} \\
\end{tabular}

\caption{Attribute transition probability by group of time interval on ML-1M, Beauty, Toys, and Sports. The x-axis is the time interval between two consecutive items, and the y-axis is the transition probability of genre/category. }\label{fig:trasition_prob_category}

\end{figure}
\subsection{Presence of User Preference Drifts} \label{appen:transition_prob}

Figure~\ref{fig:trasition_prob_category} shows the transition probability of the item's attribute by time interval group. For attribute information, we utilize the attributes of each dataset. For ML-1M, the genre that best characterizes the movie was used as the attribute. For Beauty, Toys, and Sports attributes, we utilized item categories as representative attributes. Given a user sequence $i_1 \rightarrow i_2 \rightarrow i_3$, if the genres of $i_1$ and $i_2$ are horror and comedy, then one attribute transition has occurred. If the genres of $i_2$ and $i_3$ are the same, then no attribute change has occurred, and the user sequence has experienced one attribute transition within the user sequence.
In this manner, the attribute transition probability is calculated for the entire user sequence and averaged by the time interval groups. For example, if the time interval of an item transition is 3 days, it falls in the range $1<\cdot\leq7$, and if the two items interacted at the same time, it belongs to the range $\leq 0$. Figure~\ref{fig:trasition_prob_category} indicates that as the time interval increases, the probability of attribute transition increases in the four datasets. This result indirectly means that user preference changes over time.
However, traditional linear SR models (\eg, SLIST~\cite{ChoiKLSL21}) cannot reflect user preferences that change over time because they consider the time interval between successive items with equal weight. To address this issue, our proposed TALE utilizes temporal information to effectively capture user preference drifts.

\begin{figure*}
\begin{tabular}{ccc}
\includegraphics[height=3.7cm]{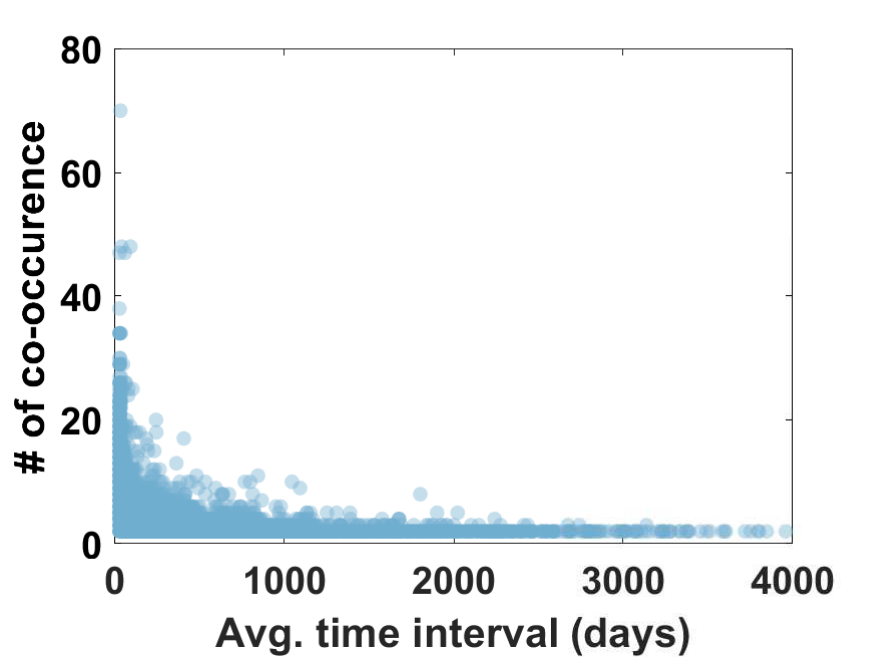} &
\includegraphics[height=3.7cm]{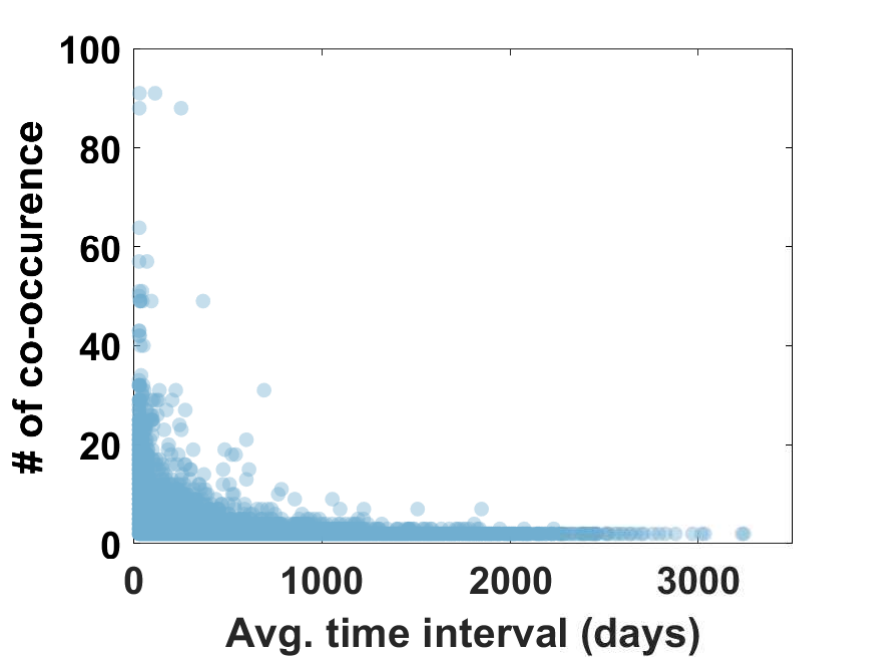} &
\includegraphics[height=3.7cm]{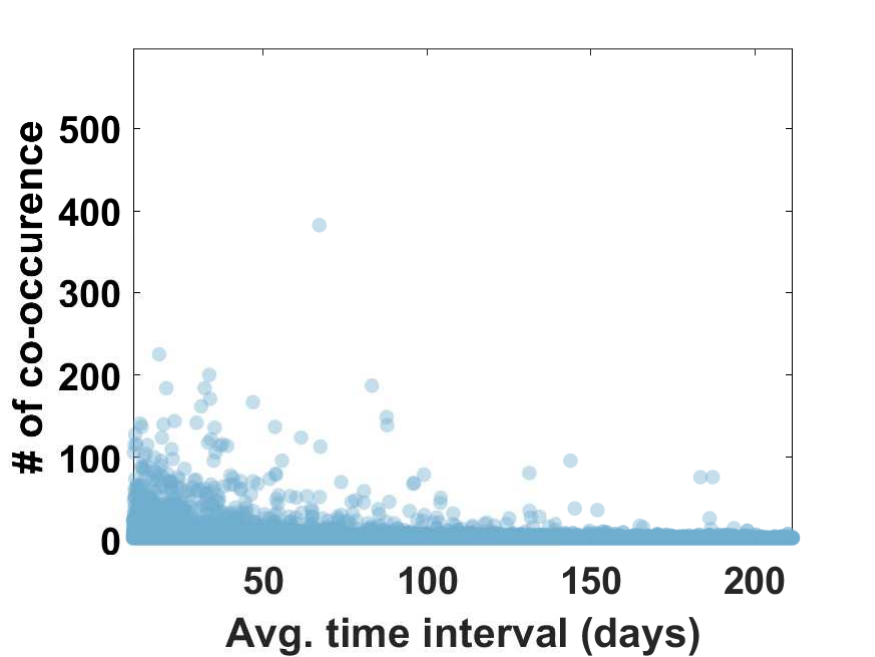} 
\\
\multicolumn{1}{c}{(a) Toys} & \multicolumn{1}{c}{(b) Sports} & \multicolumn{1}{c}{(c) Yelp}
\end{tabular}


\vspace{-2mm}
\caption{Co-occurrences between two consecutive items over average time interval on Toys, Sports, and Yelp. }\label{fig:app_intv_co_occur}

\end{figure*}

\subsection{Co-occurrences over Time Intervals} \label{appen:intv_co_occur}
Figure~\ref{fig:app_intv_co_occur} depicts the co-occurrence over average time intervals on Toys, Sports, and Yelp. They also show similar trends to ML-1M and Beauty in Figure~\ref{fig:intv_co_occur}. Note that Yelp shows a relatively uniform distribution compared to the other datasets because it is a place review dataset and sequential order is less important than others.

\begin{figure}
\begin{tabular}{cc}
\includegraphics[height=3.0cm]{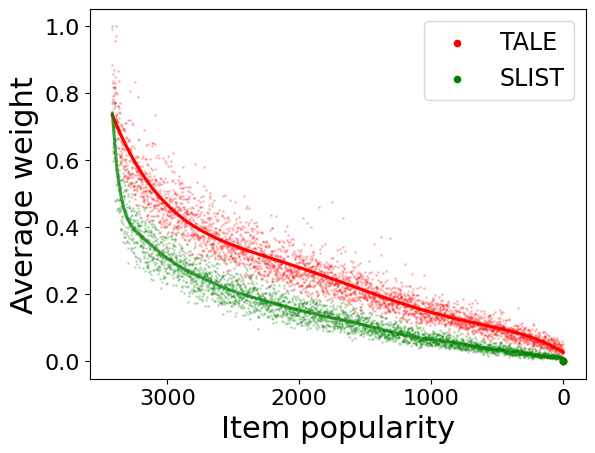} & \includegraphics[height=3.0cm]{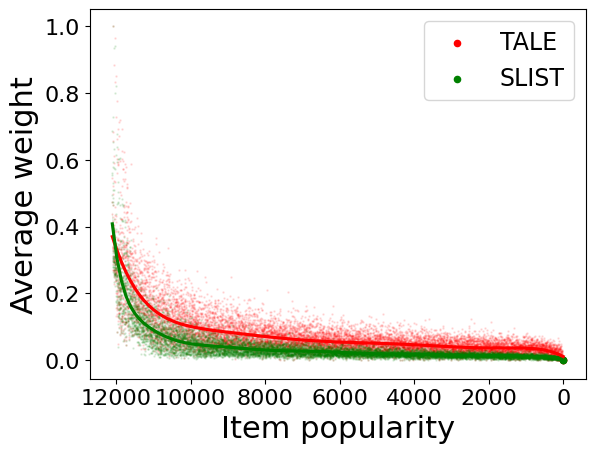} 
\\
\multicolumn{1}{c}{(a) ML-1M} & \multicolumn{1}{c}{(b) Beauty}
\end{tabular}
\vspace{-2mm}
\caption{Correlation between item popularity and the average weight for each item on ML-1M and Beauty. The x-axis is the item ID sorted by popularity, and the y-axis is the average weight for each item. }\label{fig:predict_score_}


\vspace{-2mm}
\end{figure}

\subsection{Effect of Trend-aware Normalization} \label{appen:effect_trend_norm}
Figure~\ref{fig:predict_score_} depicts the average model prediction score of each item according to item popularity.
Before describing the results, we introduce the two properties of the ideal distribution (if the model is perfectly debiased): (i) Uniform distribution. (ii) All weights with a value of 1.

We found the intriguing observations. (i) Compared to TALE, SLIST has a more skewed distribution where a few popular items have high scores, meaning that SLIST has more popularity bias. Meanwhile, TALE has a more uniform distribution because it mitigates the popularity bias. (ii) Applying trend-aware normalization elevates the performance of tail and head items by slightly increasing the weights of the overall items, as evidenced in Table~\ref{tab:ablation_performance}.

\begin{table} 
\caption{Pearson correlation coefficients between inverse average time interval matrix and the learned item-item weight matrix $\hat{\mathbf{B}}$ for SLIST and TALE on five datasets.}
\label{tab:pearson_temp}
\begin{center}
\renewcommand{\arraystretch}{1} 
\begin{tabular}{c|ccccc}
\toprule
Model    & ML-1M   & Beauty   & Toys   & Sports   & Yelp               \\
\midrule
SLIST ~\cite{ChoiKLSL21}& 0.0130 & 0.0524 & 0.0718 & 0.0325 & 0.0910 \\
TALE & \textbf{0.0375} & \textbf{0.3944} & \textbf{0.5602} & \textbf{0.5027} & \textbf{0.3885} \\

\bottomrule
\end{tabular}
\end{center}
\end{table}

\subsection{Analysis of Temporal Information} \label{appen:reflect_tempo_info}
Table~\ref{tab:pearson_temp} indicates how SLIST and TALE reflect temporal information by using the Pearson correlation coefficient. 
Inspired by the observation that the probability of user preference drifts is proportional to the average time interval (Refer to Figure~\ref{fig:trasition_prob_category}), we analyze the correlation between the average time interval of items and the learned item-item weight matrix $\hat{\mathbf{B}}$. Specifically, we first compute the average time interval between consecutive items. These values constitute the average time interval matrix of the same size as the item-item matrix. Since shorter average time intervals are generally more highly correlated, we take the element-wise inverse of the average time interval matrix. Then, we compute the Pearson correlation coefficient between this matrix and the weight matrix $\hat{\mathbf{B}}$. The larger the coefficient, the better $\hat{\mathbf{B}}$ reflects the temporal information. In Table~\ref{tab:pearson_temp}, TALE has higher coefficients than SLIST in all datasets, indicating that it effectively models user preference drifts into the item-item weight matrix.

\begin{table} 

\caption{Efficiency comparison for TALE and seven SR models on ML-1M, Beauty, and Yelp. `Train' and `Eval' mean the runtime (seconds) for training and evaluation, respectively. For neural models, the runtime was measured on GPU, and for linear models on CPU.}
\label{tab:efficiency_appen}
\vspace{-2mm}
\begin{center}
\renewcommand{\arraystretch}{1} 
\begin{tabular}{P{1.9cm}|r r|r r|r r}

\toprule
Dataset         & \multicolumn{2}{c|}{ML-1M}          & \multicolumn{2}{c|}{Beauty}         & \multicolumn{2}{c}{Yelp} \\
Model & Train & Eval & Train & Eval & Train & Eval \\
\midrule
SASRec~\cite{KangM18SASRec}           & 1,475 & 11 & 128   & 42 & 252 & 43  \\  
DuoRec~\cite{QiuHYW22}              & 6,278 & 11 & 1,308 & 42 & 2,182 & 43  \\
FEARec~\cite{DuYZQZ0LS23}           & 20,301 & 48 & 529 & 172 & 539 & 246  \\ 
BSARec~\cite{ShinCWP24}        & 2,537 & 12 & 89   & 42 & 309 & 55  \\
\midrule
TiSASRec~\cite{LiWM20}         & 5,370 & 12 & 2,093 & 43  & 1,662 & 57  \\
TCPSRec~\cite{TianLBWZ22}           & 7,853 & 11 & 421 & 42 & 1,336 & 43  \\
TiCoSeRec~\cite{DangYGJ0XSL23}          & 10,512 & 11 & 322 & 42 & 5,794 & 43  \\
\midrule
TALE           & \textbf{179} & \textbf{0.2} & \textbf{5}     & \textbf{10} & \textbf{13} & \textbf{21}  \\

\bottomrule
\end{tabular}
\vspace{-2mm}
\end{center}
\end{table}
\subsection{Efficiency Comparison} \label{appen:efficiency_comparison}
Table~\ref{tab:efficiency_appen} shows the training and evaluation time of TALE and the comparison models on three datasets. TALE has the fastest training time, 8.2x, 25.6x, and 19.4x faster than SASRec for ML-1M, Beauty, and Yelp, respectively. Since the training time was measured for a single model training, when hyperparameter tuning is performed, the neural models take a much longer training time than TALE. Among the neural models, SASRec and BSARec have the shortest training times, while the contrastive learning-based models (\ie, DuoRec, TCPSRec, and TiCoSeRec) have longer training times because of the data augmentation and the computation of contrastive loss. Looking at the efficiency of LRURec in Table~\ref{tab:efficient_models_comparison}, it has a long evaluation time compared to other neural models, which is caused by the serial processing of RNN.

\begin{figure*}
\begin{tabular}{ccc}
\includegraphics[height=3.5cm]{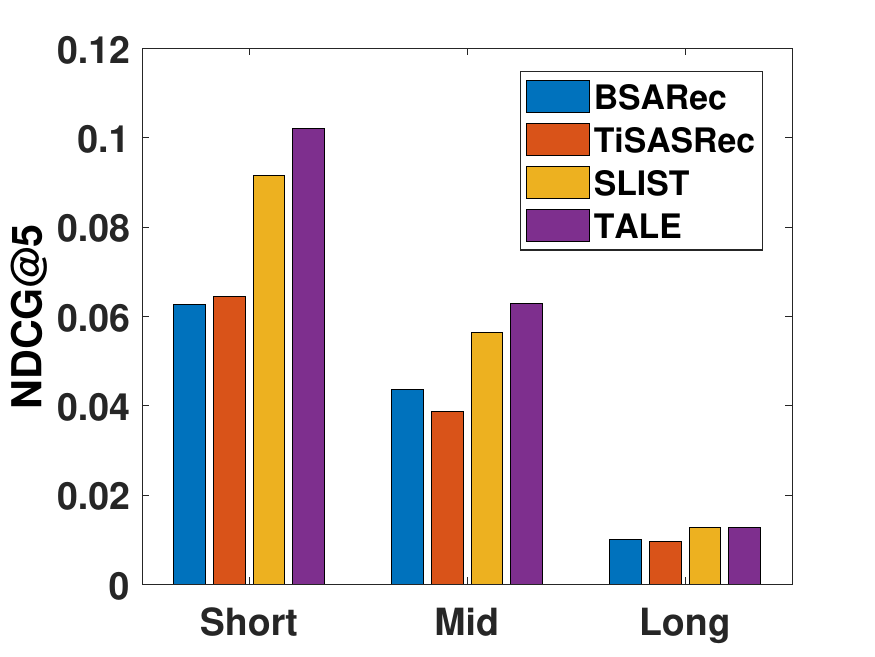} &
\includegraphics[height=3.5cm]{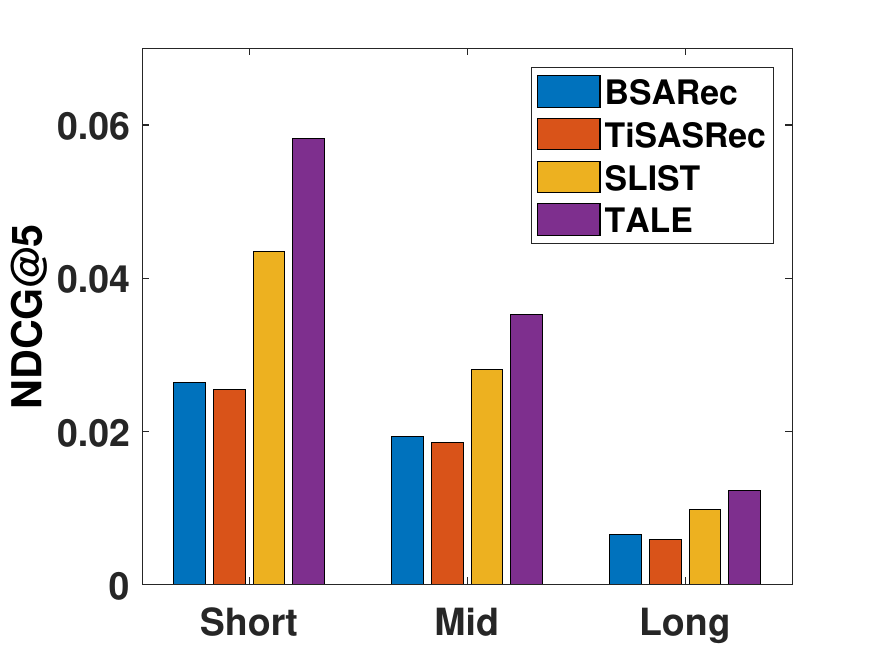} &
\includegraphics[height=3.5cm]{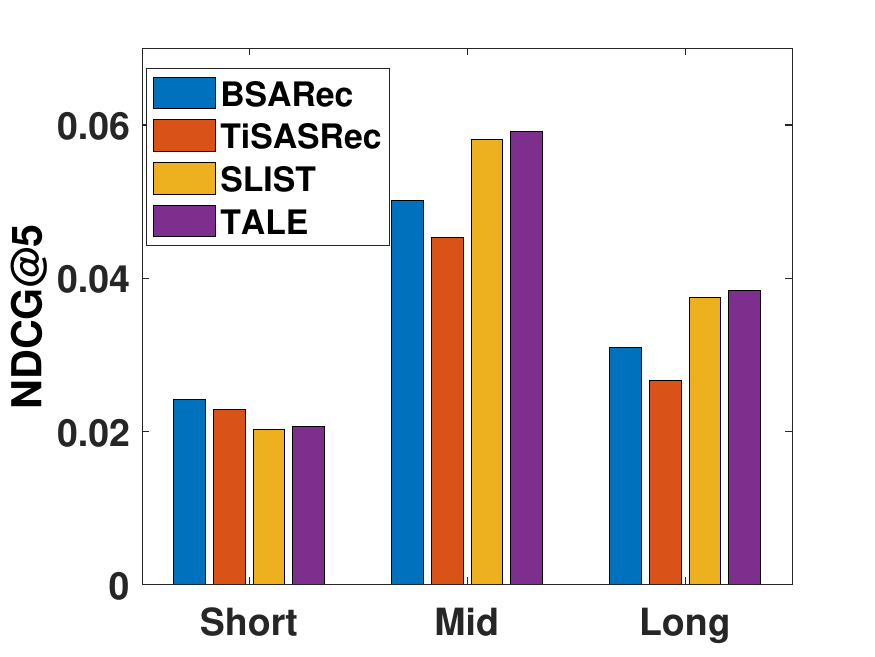}\\
\multicolumn{1}{c}{(a) Toys} & \multicolumn{1}{c}{(b) Sports} & \multicolumn{1}{c}{(c) Yelp} \\
\end{tabular}
\vspace{-3mm}
\caption{Accuracy comparison by the time interval group on Toys, Sports, and Yelp. Each metric is NDCG@5. }\label{fig:ablation_time_intv_perf_appendix}
\vspace{-1mm}
\end{figure*}
\begin{figure*}
\begin{tabular}{ccc}
\includegraphics[height=3.6cm]{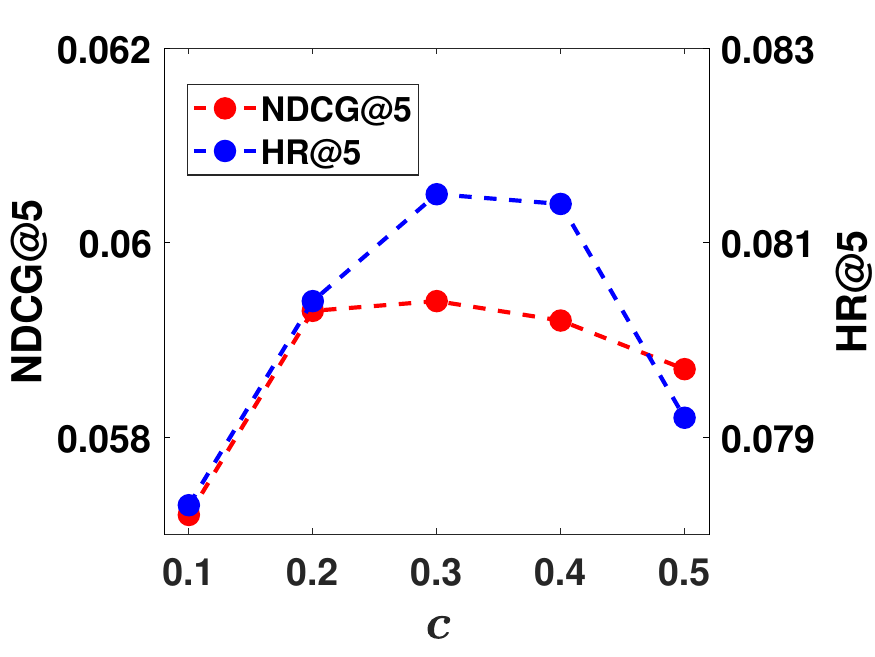} &
\includegraphics[height=3.6cm]{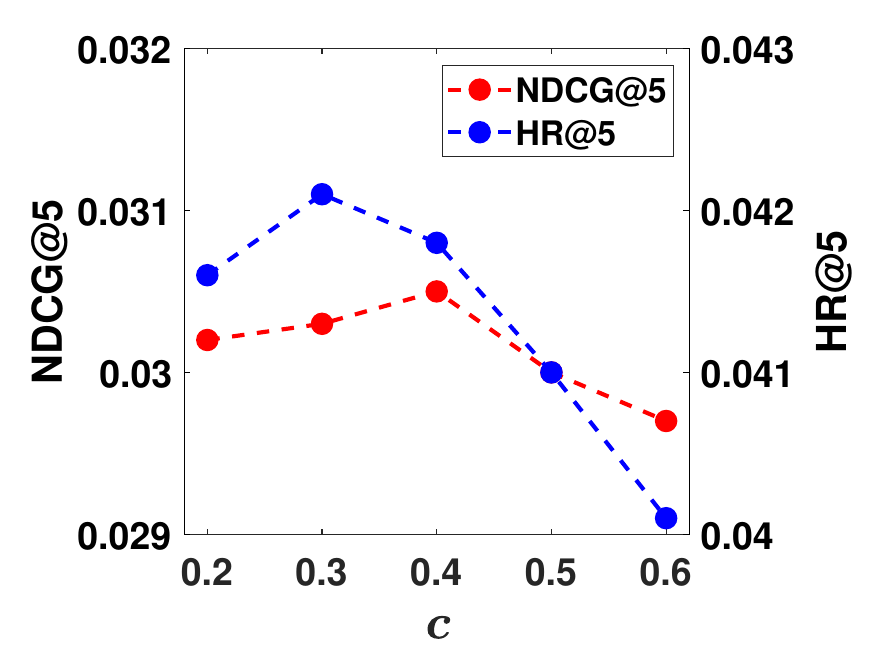} &
\includegraphics[height=3.6cm]{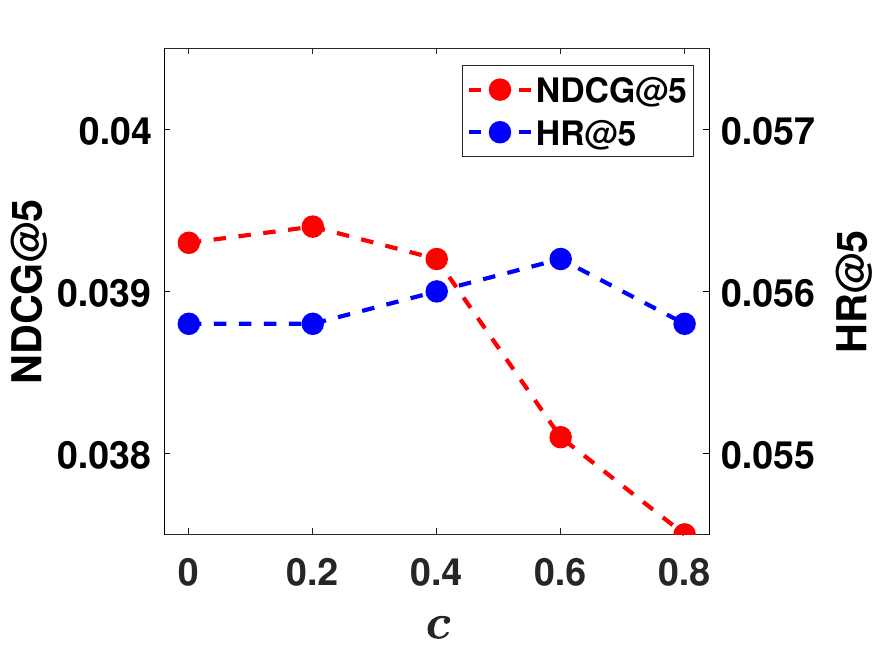}\\

\includegraphics[height=3.6cm]{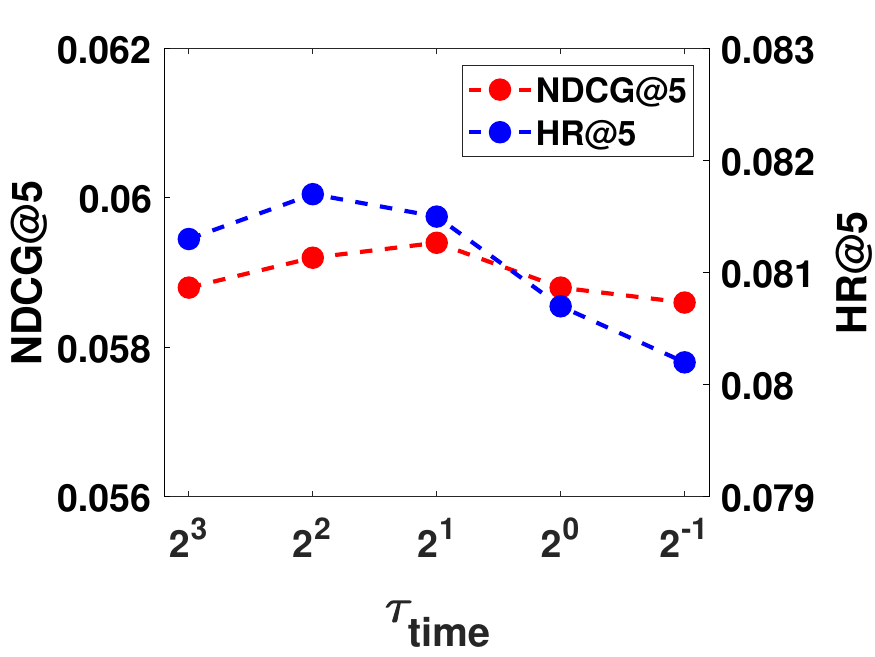} &
\includegraphics[height=3.6cm]{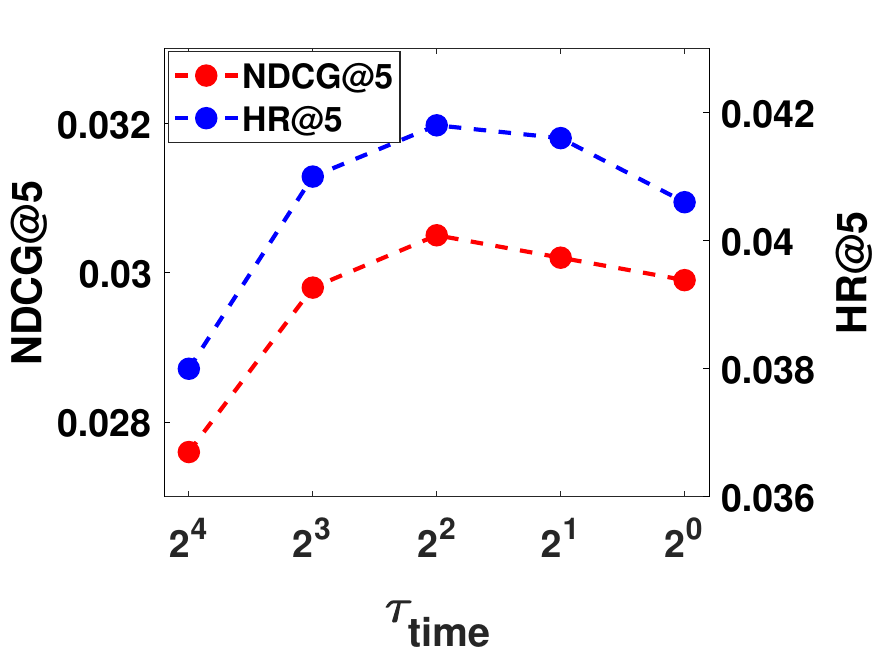} &
\includegraphics[height=3.6cm]{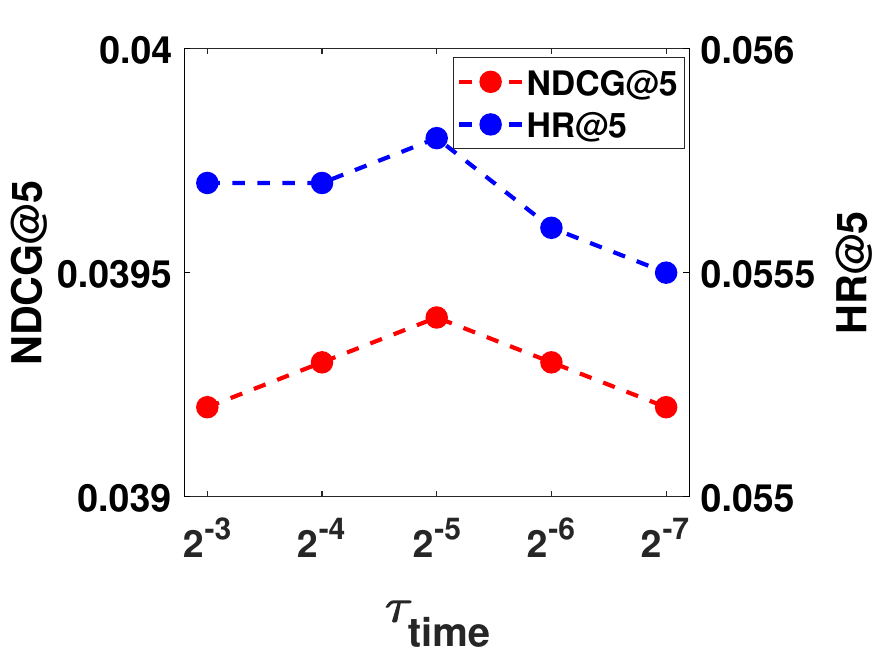}\\

\includegraphics[height=3.6cm]{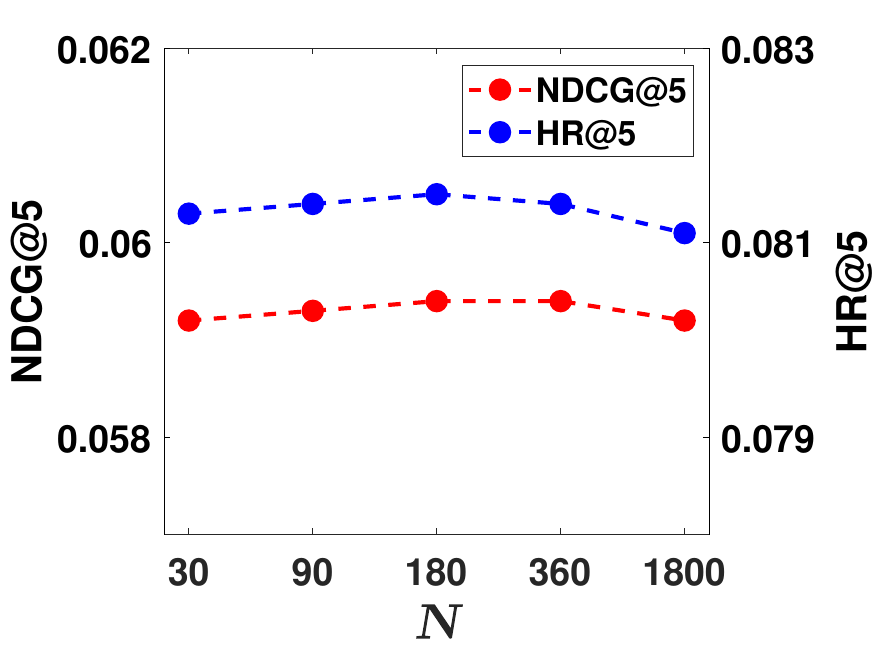} &
\includegraphics[height=3.6cm]{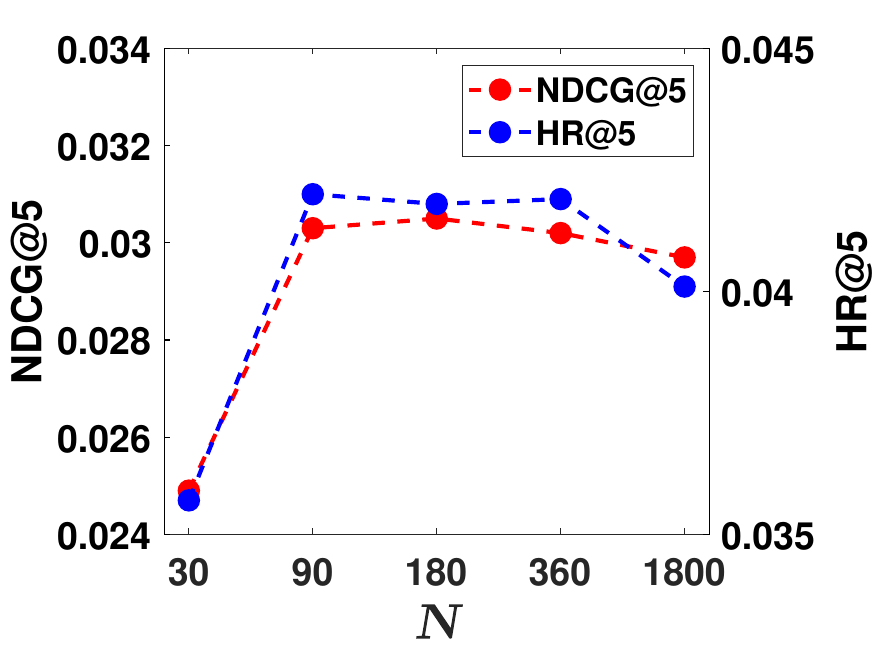} &
\includegraphics[height=3.6cm]{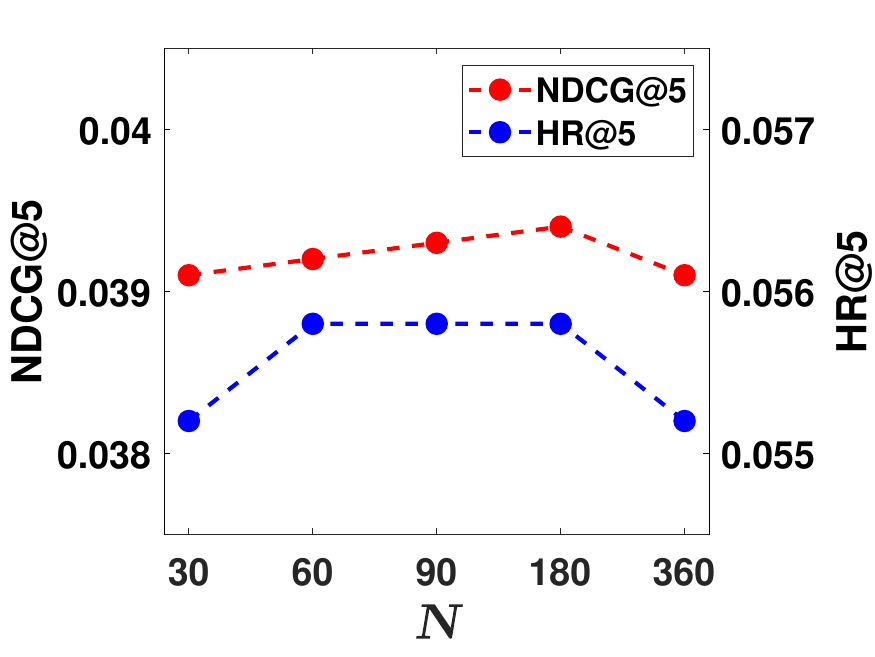}\\

\multicolumn{1}{c}{(a) Toys} & \multicolumn{1}{c}{(b) Sports} & \multicolumn{1}{c}{(c) Yelp} \\
\end{tabular}
\vspace{-2mm}
\caption{Performance of TALE over the three hyperparamters, \ie, $c$, $\tau_{\text{time}}$, and $N$, on Toys, Sports, and Yelp. }\label{fig:all_hyper_appen}
\vspace{-1mm}
\end{figure*}

\begin{table} \small
\caption{Tail and Head performance comparison on Toys, Sports, and Yelp. `Norm.' denotes the existing normalization method, \ie, Eq.~\eqref{eq:sym_norm}. Each metric is measured by NDCG@5. }
\label{tab:tail_head_appen}
\vspace{-2mm}
\begin{center}
\renewcommand{\arraystretch}{1} 
\begin{tabular}{P{1.4cm}|P{0.75cm}P{0.75cm}|P{0.75cm}P{0.75cm}|P{0.75cm}P{0.75cm}}
\toprule
Dataset         & \multicolumn{2}{c|}{Toys}          & \multicolumn{2}{c|}{Sports} & \multicolumn{2}{c}{Yelp}               \\
Model              & Tail      & Head      & Tail      & Head & Tail      & Head            \\
\midrule
SASRec & 0.0287    & 0.0431    & 0.0088    & 0.0263    & 0.0309    & 0.0330         \\
BSARec     & 0.0306    & 0.0514    & 0.0094    & 0.0297    & 0.0343    & 0.0359    \\
TiSASRec      & 0.0303    & 0.0487   & 0.0074    & 0.0281    & 0.0290    & 0.0343        \\
\midrule
SLIST     & 0.0376    & 0.0755    & 0.0100    & 0.0486    & 0.0371    & \textbf{0.0380}        \\
SLIST+Norm.               & \underline{0.0415}    & \underline{0.0767}    & \textbf{0.0146}    & \underline{0.0466}    & \underline{0.0402}    & 0.0333        \\
TALE                     & \textbf{0.0428}    & \textbf{0.0818}    & \underline{0.0133}    & \textbf{0.0512}    & \textbf{0.0415}    & \underline{0.0367}       \\

\bottomrule
\end{tabular}

\vspace{-2mm}
\end{center}
\end{table}

\subsection{Tail and Head Performance} \label{appen:tail_head_perfrom}
Table~\ref{tab:tail_head_appen} shows Tail and Head performance on Toys, Sports, and Yelp. In all three datasets, we can see that TALE outperforms other models on tail items, alleviating popularity bias. The existing normalization (\ie, Norm.) also shows superior Tail performance, indicating the debiasing effect of normalization.

\subsection{Performance over Time Intervals} \label{appen:perform_time_interval}
Figure~\ref{fig:ablation_time_intv_perf_appendix} illustrates the performance of the three datasets over the three groups as time intervals (\ie, Short, Mid, and Long). On Toys and Sports, TALE excels in all groups, similar to the Beauty dataset, while the Yelp dataset performs well in Mid and Long. The consistent performance in Mid and Long suggests that TALE successfully captures the relation between items with long-time intervals. Meanwhile, the performance of the Yelp dataset for Short is lower than for Mid and Long because users do not consume similar items in a row, which is a characteristic of the Yelp dataset. For example, a user visiting a Chinese restaurant does not immediately visit another Chinese restaurant. 


\subsection{Hyperparameter Sensitivity} \label{appen:hyper_sensi}
Figure~\ref{fig:all_hyper_appen} shows the performance of TALE on Toys, Sports, and Yelp according to the three hyperparameters (\ie, $c$, $\tau_{\text{time}}$, and $N$). Based on the dataset statistics, we divide them into three groups. \emph{Group1}: (Beauty, Toys, and Sports), \emph{Group2}: Yelp, and \emph{Group3}: ML-1M. The average time interval and optimal $\tau_{\text{time}}$ become smaller in the order of Group 1, 2, and 3. This is because it is natural to give a weaker time decay for shorter average time intervals (The numerator and denominator become similar in scale.). On Toys, Sports, and Yelp, optimal values of $c$ are 0.3, 0.4, and 0.2, respectively, indicating that long-time dependency varied depending on the dataset characteristics. The three datasets achieve optimal performance when the window size $N$ for trend-aware normalization is 180.

\end{document}